\documentclass[9pt]{article}
\usepackage[margin=1in]{geometry}                
\usepackage[parfill]{parskip}    
\usepackage{multicol}
\usepackage{authblk}
\usepackage{mathtools}
\usepackage{caption}
\usepackage{subcaption}
\usepackage{graphicx}
\usepackage{float}
\usepackage[numbers]{natbib}
\usepackage{amssymb}
\usepackage{tensor}
\usepackage{xcolor}
\usepackage[english]{babel}
\usepackage{epstopdf} 
\usepackage{amsmath}
\numberwithin{equation}{section}
\usepackage{mathrsfs}
\usepackage{varioref}
\usepackage{hyperref} 
\usepackage{textcomp}
\usepackage{xcolor}
\usepackage{ifsym}
\DeclareMathOperator{\arctanh}{ArcTanh}

\newcommand{\appendixpage}

\hypersetup{
    pdfstartview={FitH},    
    colorlinks=true,       
    linkcolor=blue,          
    citecolor=blue,   
    filecolor=blue,  
    urlcolor=blue       
}

\DeclareFontFamily{OT1}{pzc}{}
\DeclareFontShape{OT1}{pzc}{m}{it}%
             {<-> s * [0.900] pzcmi7t}{}
\DeclareMathAlphabet{\mathscr}{OT1}{pzc}%
                                 {m}{it}

\begin{document}
\definecolor{orange}{rgb}{0.9,0.45,0}
\definecolor{applegreen}{rgb}{0.655, 0.251, 0.00}
\definecolor{blue}{rgb}{0.0,0.0,1.0}
\definecolor{red}{rgb}{1.0,0.0,1.0}
\newcommand{\spl}[1]{{\textcolor{olive}{[SP: #1]}}}

\newcommand{\dan}[1]{{\textcolor{applegreen}{[comment DM: #1]}}}
\newcommand{\ser}[1]{{\textcolor{cyan}{#1}}}
\newcommand{\mc}[1]{{\textcolor{blue}{[#1]}}}
\def\ren{\ensuremath{^{\mbox{\tiny ren}}}_{\mbox{\tiny os}}}
\def\extC{\ensuremath{^{\mbox{\tiny ext}}}_{\mbox{\tiny C}}}
\def\extNC{\ensuremath{^{\mbox{\tiny ext}}}_{\mbox{\tiny NC}}}
\def\nNC{\ensuremath{^{\mbox{\tiny near}}}_{\mbox{\tiny NC}}}
\def\con{\ensuremath{_{\mbox{\tiny C}}}}
\def\h{\ensuremath{_{\mbox{\tiny H}}}}
\def\nh{\ensuremath{_{\mbox{\tiny NCNENH}}}}
\def\os{\ensuremath{_{\mbox{\tiny os}}}}
\def\Ncon{\ensuremath{_{\mbox{\tiny NC}}}}
\def\effnc{\ensuremath{^{\mbox{\tiny NC}}}_{\mbox{\tiny eff}}}
\def\effc{\ensuremath{^{\mbox{\tiny C}}}_{\mbox{\tiny eff}}}

\title{\bf{Noncommutative AdS black hole and the IR holographic superconductor}}
\author[1]{\thanks{\href{mailto: mdelacruz@ifbuap.buap.mx}{ 
{\color{black}Corresponding author} mdelacruz@ifuap.buap.mx}}Manuel de la Cruz-López}
\author[1]{\thanks{\href{mailto: aherrera@ifbuap.buap.mx}{aherrera@ifuap.buap.mx}}Alfredo Herrera-Aguilar}
\author[2,3]{\thanks{\href{mailto: daniel.martinez.carbajal@correo.nucleares.unam.mx}{daniel.martinez.carbajal@correo.nucleares.unam.mx}}Daniel Martínez-Carbajal}
\author[3]{\thanks{\href{mailto: sergio.patino@correo.nucleares.unam.mx}{sergio.patino@correo.nucleares.unam.mx}}Sergio Patiño-López}

\affil[1]{\begin{small}Instituto de Física, Benemérita Universidad Autónoma de Puebla, Apdo. Postal J-48, CP 72570, Puebla, México\end{small}}
\affil[2]{\begin{small}TecNM-Tecnológico de Estudios Superiores del Oriente del Estado de México, Paraje San Isidro, Barrio de Tecamachalco,
Apdo. Postal 56400, Estado de México, México\end{small}}
\affil[3]{\begin{small}Instituto de Ciencias Nucleares, Universidad Nacional Autónoma de México, Apdo. Postal 70-543, CDMX 04510, México\end{small}}

\date{}                                           

\maketitle

\begin{abstract}
We construct a noncommutative (NC) AdS$_4$-charged black hole with a planar horizon topology. The NC effects of this geometry are captured by a Gaussian distribution of black hole mass codified in a fluid-like energy-momentum tensor. A natural bound in radial coordinate is established, below which the scalar curvature changes its sign and defines a NC cutoff that embeds the point singularity.  We study in detail the thermodynamic structure of this scenario, finding a well-defined black hole mass and an analytic criterion for its stability. Focusing on the AdS$_{2}$ structure near the horizon, we find a novel effective curvature radius with dependency on the NC cutoff. These results motivate us to explore the holographic superconducting system in terms of the nearness from the cutoff. The behavior of the magnetic field in the deep IR geometry is studied and we found semi-analytical novel expressions for the upper critical magnetic fields of a dual type-II superconductor in the canonical and grand canonical ensembles. The condensation in the form of hair is studied in terms of the bound states of the associated Schrödinger potential of the scalar field, interpreted as the dual to the density of Cooper pairs. The NC effects increase the hair formation due to a steeper AdS$_2$ throat comparable to the commutative case. Finally, we obtain the effective IR scalar field equation on the near horizon and near extremal NC Schwarzschild AdS$_2$ geometry and confirm that NC effects promote bound states that the commutative version forbids.

\end{abstract}
\section{Introduction and summary}
The program of holographic superconductivity, inaugurated by \cite{Hartnoll:2008vx, Hartnoll:2008kx} and based on the celebrated AdS/CFT correspondence \cite{Maldacena:1997re}, has been subject to intense study. It has been shown to reproduce very well the qualitative and quantitative behavior of thermodynamic quantities associated with superconductivity, described effectively with the Ginzburg-Landau mean field theory under strong coupling. Holographic superconductivity operates in the large N limit, with weak curvature and correlation functions given by the partition function of classical gravity, the so-called bottom-up holography \cite{zaanen_liu_sun_schalm_2015}.

As long as the scenario for studying Condensed Matter Systems within gravity has an asymptotically anti-de Sitter (or Lifshitz) structure, the AdS/CFT correspondence ensures the framework on which the dual system can be described. The most natural and successful scenario is the Schwarzschild-AdS black hole on which the Hawking temperature is the temperature of the dual system and scales linearly with the horizon radius. Implementing the so-called decoupling limit \cite{Hartnoll:2008vx}, the system gives rise to the Abelian-Higgs theory with a global $U(1)$ conserved current. From the early days of the implementation of the above configuration, many attempts to improve the modeling of the properties of the critical magnetic field, thermal condensate (dual to the density of superconducting Cooper pairs), and conductivity, have been realized with success. To name a few: the effects of black hole rotational parameters are traced in qualitative changes of the droplet and vortex solutions \cite{PhysRevD.106.L081902, Herrera-Mendoza:2024vfj} (also see \cite{Nakano:2008xc} for the numerical proof of condensation in a magnetically charged background), and the higher curvature corrections within Einstein-Gauss-Bonnet theory, that result in an effective AdS curvature radius and crossing lines of the thermal magnetic field shape \cite{Gregory:2009fj}. Within the Ginzburg-Landau theory, the theoretical behavior of Type-II superconductors seems to be very robust, regardless of their material constitution. Therefore, improvements in the holographic Abelian-Higgs framework must be concomitant with extensions of the Ginzburg-Landau theory.

An essential aspect of the Abelian-Higgs holographic theory on spacetimes sourced by neutral black holes is the interaction between gauge fields and scalar fields. In \cite{Gubser:2008px} and \cite{Hertog:2006rr} it was proven that their interaction is crucial for hair formation (the scalar field mass acquires contribution from the $t$-component of the gauge field). A further development, considering the full backreaction between gravity and the Abelian-Higgs sector, shows interesting physics: the hair formation arises independent of the charge $q$ of the scalar field. Moreover, the hair condensate can be sustained by the gravitational throat of the near horizon geometry which has topology AdS$_{2}\times \mathbb{R}^{2}$, even with a small electric charge $q$ (hence goes beyond the decoupling limit $q\mapsto\infty$) \cite{Hartnoll:2008kx}. Additionally, the interesting phenomena of Abrikosov lattice have been realized by using holographic setups, for instance,  considering nonbackreacted matter in \cite{Montull:2009fe, Maeda:2009vf, Albash:2009iq, Adams:2012pj, Srivastav:2023qof, Xia:2019eje}  and with full backreaction in \cite{Donos:2020viz}. 

Charged dyonic black holes have been known for a long time \cite{Romans:1991nq}. In this case, the dependency of temperature on the horizon radius becomes more intricate. A perturbative scalar field over the Einstein-Maxwell-AdS bulk constitutes one of the seeds of scalar hair formation driven by instabilities of the background \cite{Gubser:2008px}. In this system, there is no scalar contribution to the energy-momentum tensor in the Einstein equations, nevertheless, it has important developments, namely, it was used to study the Hall current \cite{Hartnoll:2007ai}, the low-frequency conductivity \cite{Horowitz:2009ij}, and the IR quantum criticality of the strange metals \cite{Faulkner:2009wj}; to name a few important contributions. Prominent for this work, is the analysis carried out in the second seminal work \cite{Hartnoll:2008kx}, in which the onset of holographic superconductivity with a perturbative scalar field, is shown in the dyonic background, exhibiting the formation of the non-trivial magnetic field profile in the canonical ensemble. The near extremal limit, close to zero temperature, is analyzed in \cite{Albash:2008eh}, demonstrating the existence of a droplet solution with Hermite polynomial behavior. In these latter works, the scalar field was considered to be purely real.

In this work, we construct a noncommutative (NC) asymptotically AdS dyonic black hole with a planar horizon topology and then consider the scalar field as a perturbation using a suitable decoupling limit from the Einstein-Maxwell-Scalar theory. The Maxwell electric and magnetic components maintain their commutative (C) character; therefore, the NC effects are controlled solely by the Nicolini energy-momentum tensor with fluid-like entries \cite{Nicolini:2008aj}. It is worth mentioning that our NC black hole construction does not solve the \emph{problem} of the essential singularity at the origin, as can be seen from the Kretschmann scalar. In this work, we are not interested in addressing the principal incentive of NC theories, namely, the complete regular solution up to a scale given by the NC parameter $\theta$.

Therefore, our configuration physically represents an NC (discretized spacetime) confining box scenario where a perturbative, charged, and commutative scalar field, \emph{feels} the NC geometry and the C-electromagnetic charges. Despite this configuration is a particular case of more general NC constructions, for instance, NC Reissner-Nordström asymptotically flat \cite{Ansoldi:2006vg}, holographic superconductor using a background with NC distribution of electromagnetic charges \cite{Pramanik:2015eka, Pramanik:2014mya} and AdS Einstein-Born-Infeld Electrodynamics with NC contributions \cite{Maceda:2019woa}; to name a few, our choice to study this spacetime has three principal motivations:
\begin{itemize}
    \item Our setup allows us to get access to the deep IR AdS$_2\times\mathbb{R}^{2}$ geometry near the horizon and the near-extremal conditions in a closed-form, controlling by the way, the NC effects with the use of a defined \emph{nearness parameter} that naturally emerges from the solution of the Einstein equations and the structure of the Ricci scalar curvature. This parameter has a minimal value below which the AdS global curvature can change its negative character, therefore, it constitutes a cutoff that avoids this pathological behavior since AdS signature is mandatory in holography. By writing the nearness parameter (from the cutoff) in terms of the horizon radius, it acquires the physical interpretation of the size of the event horizon, relative to the NC scale\footnote{The NC scale is the one in which the spacetime is discretized in cells of size given by the Planck scale \cite{Nicolini:2008aj}.}. See sections \ref{decoupling}, \ref{alphasubsec}. 
    \item We found a consistent black hole mass using holographic techniques \cite{Balasubramanian:1999re} and a novel criterion to explore the instabilities of the bulk, in terms of an equation of state. We found that the possible instabilities are closely related to the nearness parameter, constituting another evidence of the consistency of the solution. See section \ref{thermo}.
    \item Although our configuration is simpler compared to the Abelian-Higgs model in the sense that we can not obtain the thermal behavior of the condensate, we can certainly exhibit strong evidence of its existence, i.e., the hair formation in the near horizon geometry. By constructing the Schrödinger-like potential of the scalar field, we look for bound states and give them the interpretation of a density of Cooper pairs that accounts for the superconductivity density (non-zero vacuum expectation value VEV), dual to the boundary values of the scalar field. Moreover, a suitable change of coordinates allows us to get access to the near-horizon and near-extremal deep IR geometry. The IR effective potential adds more evidence for the condensate (hair) existence. A novel result of this construction is that NC effects act in favor of hair formation due to a stronger effective AdS$_{2}$ curvature relative to the C-dyonic solution and, for allowed bound states of the C-version, the potential wells become deeper when the NC effects are turned on. As a final remark of this third motivation, the nearness parameter also arises in the effective AdS$_2$ curvature radius with a physically clearer interpretation: below the cutoff of the nearness parameter, the effective AdS$_{2}$ curvature grows without bound, turning the geometry less curved, in conflict with the general knowledge about the near horizon geometry of the AdS charged black hole solutions. See sections \ref{section3}, \ref{Spotential1}, \ref{Spotential2}.
\end{itemize}
 
Since our setup allows us to holographically describe the thermal behavior of the \emph{upper critical magnetic} field of a type-II superconductor in the canonical and grand canonical ensemble \cite{book:17888}, we study the NC effects on the shape of the magnetic field in virtue of the nearness parameter. Consistent with the Schrödinger potential that improves the density of Cooper pairs, the magnetic field \emph{strengthens} their behavior relative to the commutative framework and, when we take the nearness parameter far away from the cutoff,  the magnetic field reduces continuously to the commutative case. Section \ref{upperCmagnetic}.


The commutative dyonic black hole and holographic superconductivity have been developed for some of the current authors \cite{refId0}. In this work, we take seriously the establishment of a novel analytical limit on which all the NC effects vanish, leaving the entire analyzed quantities exactly equal to the aforementioned work. As far as the authors know, these results have not been previously addressed in the literature.

\section{The setup}\label{setup}
Consider the Einstein-Maxwell-Scalar theory (EMS) with negative cosmological constant $\Lambda$ in $(3+1)$-dimensions
\begin{equation}\label{Action}
    S=\dfrac{1}{\kappa^2}\int d^{4}x\sqrt{-g}\left[R-2\Lambda-\kappa^{2}\,\mathcal{I}-\dfrac{\epsilon_{1}^{2} L^2}{4}F^2-\epsilon_{2}^2 L^2\left(\vert D\psi\vert^2+m^2\vert\psi\vert^2\right)\right],
\end{equation}
where $\kappa^{2}=8\pi G$, $R$ stands for the Ricci curvature invariant, $F$ is the Maxwell field strength tensor and $\psi$ corresponds to the scalar field with mass $m$ and $D_{a}=\nabla_{a}-i q A_{a}$ is the covariant-gauge derivative. Also, the \emph{NC energy-momentum} tensor is defined via the invariant $\mathcal{I}$ as follows
\begin{equation}\label{nicolinitensor}
    \mathcal{I}=g^{ab}\mathcal{I}_{ab}, \ \ \ \ \mathcal{I}_{ab}\equiv\left(T^{\theta}_{ab}-\dfrac{1}{2}T^{\theta} g_{ab}\right).
\end{equation}
It is straightforward to show that, for the spacetime metrics with AdS asymptotics and planar horizon topology,
\begin{equation}\label{metric0}
    ds^{2}=-U(r)dt^2+\dfrac{r^2}{L^2}\left(dx^{2}+dy^{2}\right)+\dfrac{dr^2}{U(r)},
\end{equation}
where $U(r)$ is the metric potential, the energy-momentum tensor (\ref{nicolinitensor}) satisfies the divergenceless condition $\nabla^{a}T_{ab}=0$. Also, it should be noted the appearance of the coupling $\kappa^2$ controls the NC effects on the curvature in virtue of the invariant $\mathcal{I}$ with their associated energy-momentum tensor $T^{\theta}_{ab}$. Also, the couplings $\epsilon_{i}^{2}\equiv\kappa^2/g_{m}$ ($\kappa^2/g_{s}$) control the interaction between Ricci curvature and Maxwell field strength and scalar field, respectively. Besides, the NC effects are captured by the functions \cite{Nicolini:2008aj}
\begin{equation}
    \rho_{\theta}(r)=-\dfrac{1}{2\sqrt{\pi}}\dfrac{L^2}{\mathcal{V}_{2}}\dfrac{m}{\theta^{3/2}}\exp{\left(-r^2/4\theta\right)}, \ \ \ \ p_{\perp}=-\dfrac{1}{2r}\left(r^2\rho_{\theta}\right)_{,r}, \quad p_{r}=-\rho_{\theta},
\end{equation}
where $m$ is the black hole \emph{bare} mass that will be identified with the physical mass once we compute the Noether charges. Concerning the energy-momentum tensor, the above configuration corresponds to a fluid-like distribution
\begin{equation}\label{NicoloniDist}
 \left[{\tensor{T}{^a_b}}\right]=\textit{diag}\left(-\rho_{\theta},p_{\perp},p_{\perp},p_{r}\right).
\end{equation}
A comment is in order here. Since we look for metrics of the form (\ref{metric0}), the functional dependency of the function $\rho_{\theta}$ has changed relative to the standard definitions in NC black holes (Schwarzschild black hole, NC charged and DBI \cite{Nicolini:2008aj, Ansoldi:2006vg, Pramanik:2015eka, Maceda:2019woa}, for instance). The reason for doing this is that it follows straightforwardly from Gauss law that the total energy of the background yields
\begin{equation}
    m=\int_{\Sigma_{t}}d\sigma^{a}\tensor{T}{^0_a},
\end{equation}
in geometries such as AdS. Hence $\mathcal{V}_{2}$ corresponds to the $2$-dimensional volume in the transversal coordinates (\ref{metric0}) and $L$ is the \emph{bare} AdS curvature radius. Besides, the trace of the matrix arrangement (\ref{NicoloniDist}) gives
\begin{equation}
T^{\theta}=2\left(p_{\perp}-\rho_{\theta}\right)=-\mathcal{I},
\end{equation}
ensuring that the NC part of the action is captured by the tensor $\mathcal{I}_{ab}$ and physically describes a self-gravitational system whose pressure acts against the gravitational pulling inward. 
\subsection{Decoupling limit and the NC dyonic background}\label{decoupling}
\begin{figure}[h]
    \centering
    \includegraphics[width=0.5\textwidth]{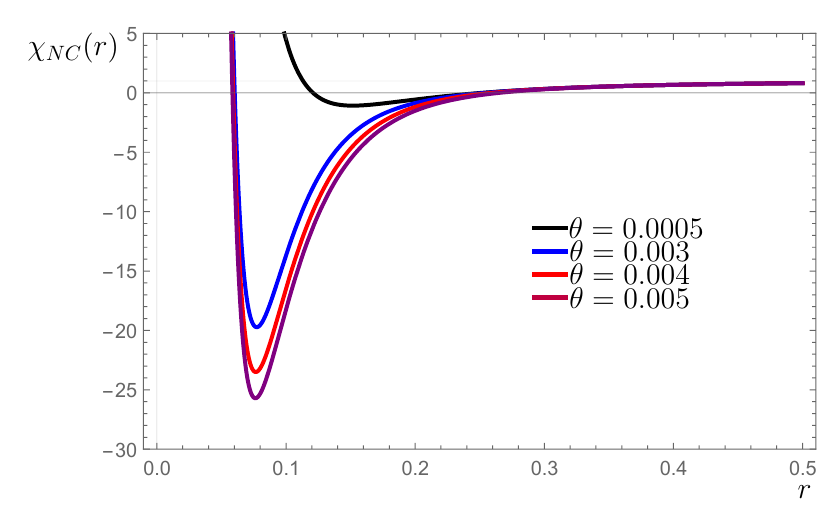}
    \caption{\begin{small}The blackening factor $\chi\Ncon (r)$ for different values of the noncommutative parameter $\theta$. It shows a continuous transition from the solution with two horizons, an extremal case, and no horizon structure. As $r\mapsto\infty$, the function $\chi\Ncon(r)\mapsto 1$, as it should be. \end{small}}
    \label{fig:outerh}
\end{figure}

Identifying $g_{s}$ with the charge $q$ of the scalar field and performing the rescalings $\psi\mapsto\psi/q$, $A\mapsto A/q$, enable us to consider the limit $q\mapsto\infty$ while $g_{m}\mapsto 0$. This procedure implies that the effect of the scalar field sector in the action (\ref{Action}), can be neglected. This configuration considers the scalar field as a perturbation over the geometry supported by the electromagnetic fields and the noncommutative mass distribution, i.e., a noncommutative dyonic AdS-black hole (NCDAdS). Therefore, the relevant equations of motion that entail the action (\ref{Action}) in the decoupled limit, read
\begin{align}
    R_{ab}-\Lambda g_{ab} &=\kappa^{2}\mathcal{I}_{ab} -\dfrac{L^2 \epsilon_{1}^{2}}{8}F^2 g_{ab}+\dfrac{L^{2}\epsilon_{1}^{2}}{2}\tensor{F}{_a_c}\tensor{F}{_b^c}\ ,\\
    \nabla^{a}F_{ab}&=0 \ .
\end{align}
Exploiting the $U(1)$-gauge invariance of Electrodynamics, we consider a magnetic charge in the London gauge
\begin{equation}
    A=A_{t}(r)dt+Bx_{1}dx_{2}.
\end{equation} 
Under these ansätze, the following system arises
\begin{subequations}\label{EOMS}
\begin{align}
    2U'(r)+rU''(r)-\left(-\Lambda-\kappa^{2}p_{\perp}(r)+\dfrac{\epsilon_{1}^{2}L^{2}}{4}\left(A_{t}'^{2}+\dfrac{L^{4}}{r^{4}}B^{2}\right)\right)2r &=0\ ,\label{R00}\\
    rU'(r)+U(r)+\left(\Lambda+\kappa^{2} \rho_{\theta}(r)+\dfrac{\epsilon_{1}^{2}L^{2}}{4} A_{t}'^{2}\right)r^{2}+\dfrac{\epsilon_{1}^{2}L^{6}}{4r^{2}}B^{2} &=0\ ,\label{constriction}\\
    A_{t}''+\dfrac{2}{r}A_{t}' &=0 \ ,\label{Maxt}\\
    \partial^{2}_{x_{1}}A_{x_{2}} &=0 \ ,\label{MaxX}
\end{align}
\end{subequations}
being $'\equiv\frac{\partial}{\partial r}$. The above system admits the analytical solution\footnote{In \cite{Ansoldi:2006vg,Pramanik:2015eka,Pramanik:2014mya} it was shown that with NC charges, there is also an analytical solution. Therefore, our construction constitutes a particular case of these works.}
\begin{subequations}\label{Sol1}
\begin{align}
    A_{t}(r) &= c_{2}-\frac{c_1}{r},\label{maxsol}\\
    U(r) &=-\dfrac{\Lambda}{3}r^{2}-\dfrac{c_3}{r}+\dfrac{\epsilon_{1}^2L^2}{4r^2}\left(c_{1}^2+L^4 B^2\right)+\dfrac{\kappa^{2}L^{2}}{\mathcal{V}_{2}}\dfrac{m}{r}\left(-\dfrac{r}{\sqrt{\pi\theta}}e^{-r^2/4\theta}+\text{erf}\left(\dfrac{r}{2\sqrt{\theta}}\right)\right)+c_{4}.\label{MetricSol}
\end{align}
\end{subequations}
Substituting (\ref{R00}) into (\ref{constriction}) we get $c_{4}=0$ identically. Since we shall explore the NC effects on the holographic superconductor, we can associate $c_{1}$ with the charge density $Q$ and $c_{2}$ with the chemical potential $\mu$ at the AdS boundary (i.e., the VEV and source in the dual theory, respectively). Finally, using the boundary stress tensor method \cite{Balasubramanian:1999re}, the total mass/energy of the system reads
\begin{equation}
   \text{Q}_{\zeta}=\int d^{2}x\sqrt{\sigma}\zeta^{\mu}T_{\mu\nu}k^{\nu}=m,
\end{equation}
where the integration is performed over the $2$-dimensional transversal coordinates with metric $\sigma_{ij}$ and $\zeta^{\mu}$, $k^{\nu}$ correspond to the time-like vector orthogonal to the $t$-constant surface and the Killing vector for the time translations, respectively. To go further in the analysis of the solution of the system (\ref{EOMS}), let us denote
\begin{equation}\label{reparam1}
    \mathcal{Q}\equiv\dfrac{L^{2}\epsilon_{1}}{2} Q, \ \ \ \ \mathcal{B}\equiv \dfrac{L^4\epsilon_{1}}{2}B, \ \ \ \ \mathcal{F}\equiv \mathcal{B}^{2}+\mathcal{Q}^2,\ \ \ \ \mathcal{M}\equiv\dfrac{\kappa^{2}L^4}{\mathcal{V}_{2}}m
\end{equation}
that allows us to write the NC solution (\ref{Sol1}) as\footnote{he couplings $\epsilon_{i}$ we use to explain the decoupling limit will no longer appear hereafter.}
\begin{subequations}\label{finalSol}
\begin{align}
   A_{t}(r)&=\mu\left(1-\dfrac{Q}{\mu r}\right)\label{vectorsol},  \\
   \chi\Ncon(r)&=1-\frac{2\mathcal{M}}{r^3}\dfrac{\gamma_{l}(r;\theta)}{\sqrt{\pi}}+\dfrac{\mathcal{F}}{r^4},\label{blackeningsol}\\
   U\Ncon(r)&=\dfrac{r^2}{L^2}\chi\Ncon(r),\label{metricfunsol}
\end{align}
\end{subequations}
exhibiting the conformal factor $r^2/L^2$ at the asymptotic AdS ($r\mapsto\infty $) and the \emph{blackening factor} $\chi\Ncon(r)$. In the equation (\ref{blackeningsol}), we identify\footnote{Recall that the Gaussian error function can be defined by $\text{erf}(z)\equiv\dfrac{2}{\sqrt{\pi}}\int_{0}^{z}e^{-t^2}dt$.}
\begin{align}\label{gammaf}
   \dfrac{r}{\sqrt{\theta}}e^{-r^2/4\theta}-\sqrt{\pi}\text{erf}\left(\dfrac{r}{2\sqrt{\theta}}\right) &= -4\int_{0}^{r/2\sqrt{\theta}} z^2 e^{-z^2}dz =-2\int_{0}^{r^2/4\theta}\sqrt{t}e^{-t}dt\nonumber\\ \\
   &\equiv -\left(\sqrt{\pi}-2\gamma_{u}\left(\dfrac{3}{2},\dfrac{r^2}{4\theta}\right)\right)=-2\gamma_{l}\left(\dfrac{3}{2},\dfrac{r^2}{4\theta}\right)\nonumber,
\end{align}
being $\gamma_{u}$ ($\gamma_{l}$) the \emph{upper (lower) incomplete gamma function}, respectively \cite{10.5555/1098650}. In Fig.~\ref{fig:outerh}, we show the behavior of the blackening factor for a variety of NC parameter $\theta$ values.

To end this subsection, we stress that the exact solution (\ref{finalSol}) proves the existence of an outer horizon radius $r_{h}$, say\footnote{Although is not possible to obtain in a closed form the roots $\chi\Ncon$ in terms of the black hole parameters.}. Therefore, we can write the blackening factor in terms of this quantity. Solving for the mass in $\chi\Ncon (r_{h})=0$, and then substituting back, yields
\begin{equation}\label{bknG}
    \chi\Ncon(r) =1+\dfrac{\mathcal{F}}{r^{4}}-\dfrac{\left(r_{h}^{4}+\mathcal{F}\right)}{r_{h}r^{3}} \mathcal{G}(r),
\end{equation}
where the function $\mathcal{G}$ reads
\begin{equation}\label{Gfun}
    \mathcal{G}(r)=\dfrac{\text{erfc}\left(\frac{r}{2\sqrt{\theta}}\right)-1+\frac{r}{\sqrt{\pi\theta}}\exp\left(-r^{2}/4\theta\right)}{\text{erfc}\left(\frac{r_{h}}{2\sqrt{\theta}}\right)-1+\frac{r_{h}}{\sqrt{\pi\theta}}\exp\left(-r_{h}^{2}/4\theta\right)}=\frac{\gamma _{l}\left ( \frac{3}{2},\frac{r^{2}}{4\theta } \right )}{\gamma _{l}\left ( \frac{3}{2},\frac{r_{h}^{2}}{4\theta } \right )},
\end{equation}
satisfying $\chi\Ncon(r_{h})=0$. Also, for holographic purposes, we emphasize the relation between the chemical potential and the electric charge of the black hole 
\begin{equation}\label{qvsmu}
    \mathcal{Q}=\dfrac{L^2 \epsilon_{1}}{2}\mu r_{h},
\end{equation}
that will be important in holography since it allows us to change between canonical and grand canonical ensembles depending on which quantity is to be fixed at the boundary. From now on, we absorb the coupling $\epsilon_{1}$ into the electromagnetic charges.
\subsection{Asymptotic behaviors and the NC cutoff}\label{alphasubsec}
The principal expected limiting behavior of the solution (\ref{finalSol}) arises when the NC effects can be neglected. Since the ratio $r^2/2\theta$ captures the NC effects through the function \eqref{gammaf}, we expect the blackening factor to reduce to the commutative case when $r\gg 2\sqrt{\theta}$. Considering a series expansion around this limit on the lower gamma function
\begin{equation}
    \begin{aligned}
\gamma_{l} \left ( \frac{3}{2},\frac{r^{2}}{4\theta } \right ) &=\Gamma \left ( \frac{3}{2} \right )+\frac{r}{2\sqrt{\theta }}e^{-\frac{r^{2}}{4\theta }}\left ( 1+\frac{2\theta}{r^2}\cdots  \right) \\
 &= \frac{\sqrt{\pi }}{2}-\frac{r}{2\sqrt{\theta }}e^{-\frac{r^{2}}{4\theta }}+\cdots,
\end{aligned}
\end{equation}

it can be seen that, up to the first order in $\theta$
\begin{equation}
    U\Ncon\left ( r \right )=\frac{r^{2}}{L^{2}}-\frac{\mathcal{M}}{L^{2}r}+\frac{\mathcal{M}}{L^{2}\sqrt{\pi \theta }}e^{-\frac{r^{2}}{4\theta }} +\frac{\mathcal{F}}{L^{2}r^{2}}.
\end{equation}
The remainder NC contribution vanishes if we take the limit
\begin{equation}
   \lim_{\theta\mapsto 0} U\Ncon\left(r\right)=U\con\left(r\right),
\end{equation}
being the function $U\con$ the commutative metric potential worked out previously in the same parametrization \cite{refId0}. 

Turning our attention to the curvature scalars, we stated in the introduction section that the Kretschmann scalar $\mathcal{K}$
\begin{align}
     \pi  L^4\mathcal{K} &= 24 \pi+\dfrac{56 \pi  \mathcal{F}^2}{r^8}+\mathcal{M}^2\left[ \frac{e^{-\frac{r^2}{2 \theta }} \left(16 \theta ^2+r^4\right)}{16\theta ^5}\right.\nonumber\\
    &\quad\left.+\frac{2 \gamma_{l}(r;\theta )}{r^6}  \left(24 \gamma_{l}(r;\theta )-\frac{r^3 e^{-\frac{r^2}{4 \theta }} \left(4 \theta +r^2\right)}{\theta ^{5/2}}\right)\right]\nonumber\\
&\quad+\sqrt{\pi} \mathcal{M}\left[\frac{ e^{-\frac{r^2}{4 \theta }} \left(r^2-8 \theta \right)}{\theta ^{5/2} } +\frac{\mathcal{F}}{r^7}\left( \frac{r^3 e^{-\frac{r^2}{4 \theta }} \left(8 \theta +3 r^2\right)}{\theta ^{5/2}}-96 \gamma_{l}(r;\theta ) \right)  \right],
\end{align}
reveals the essential singularity at $r=0$ (remaining finite at the horizon). Therefore, our configuration does not address the primary motivation problem of a point singularity, with the emergence of a minimal finite size resolution at the Planck scale that accounts for NC fluctuations of the spacetime manifold \cite{Nicolini:2008aj}. Asymptotically flat Reissner-Nordsröm with NC electric charge and NC black hole matter distribution was obtained in \cite{Ansoldi:2006vg}. This work shows that, close to the \emph{smeared out} point singularity, an effective positive de-Sitter cosmological constant controls the curvature and defines the so-called NC core surrounding the healed singular point. In our configuration, it is not possible to identify such an equivalent NC core. Nevertheless, AdS/CFT theory demands an AdS structure at every point of the spacetime.

It turns out that the Ricci scalar curvature possesses an interesting feature. The negative cosmological constant (that supports the constant curvature everywhere), acquires a correction due to the NC mass distribution of the energy-momentum tensor. Using the solution of the metric function (\ref{metricfunsol}), we found
\begin{equation}
\begin{aligned}
    R&=-\frac{1}{r^{2}}\left(2U\Ncon\left(r\right)+4rU'\Ncon\left(r\right)+r^{2}U''\Ncon\left(r\right)\right)\\
    &=-\frac{12}{L^{2}}-\frac{\mathcal{M}}{4L^{2}}\left(\frac{r^{2}-8\theta}{\sqrt{\pi}\theta^{5/2}}\right)e^{-\frac{r^{2}}{4\theta}}.
    \label{ricciNC}
\end{aligned}
\end{equation}
The effect of \emph{spreading out} the mass of the black hole vanishes as long as we take $\theta\mapsto 0$, recovering the curvature of the dyonic commutative black hole $R=-12/L^2$. The reader should note that, from (\ref{ricciNC}), it is mandatory to have $r^2\geq 8\theta$ to avoid potentially dangerous positive corrections to the negative curvature\footnote{Furthermore, we do not want NC corrections to make the curvature less negative than $-12/L^2$, see next lines in the main text.}. Therefore, we can define a radial \emph{NC cutoff} above which the negative curvature maintains its AdS signature. Since the scalar curvature is insensitive to the horizon position, the above condition invites us to consider the \emph{nearness parameter} from the NC cutoff
\begin{equation}\label{thebound0}
    \alpha\equiv \dfrac{r_{h}}{2\sqrt{2\theta}}\geq 1,
\end{equation}
such that, $\alpha=1$ defines the cutoff. Note that, for this value, the NC corrections to the scalar curvature vanish\footnote{From the first principles regarding the nature of the discretized spacetime, the $\theta$ parameter is of the order of the (square) Planck length.}. It turns out that, for values slightly greater than one, the NC effects become relevant, and conversely, if $\alpha\gg 1$ we expect that the NC corrections may be neglected. 

\begin{figure}[h]
    \centering
    \includegraphics[width=0.5\textwidth]{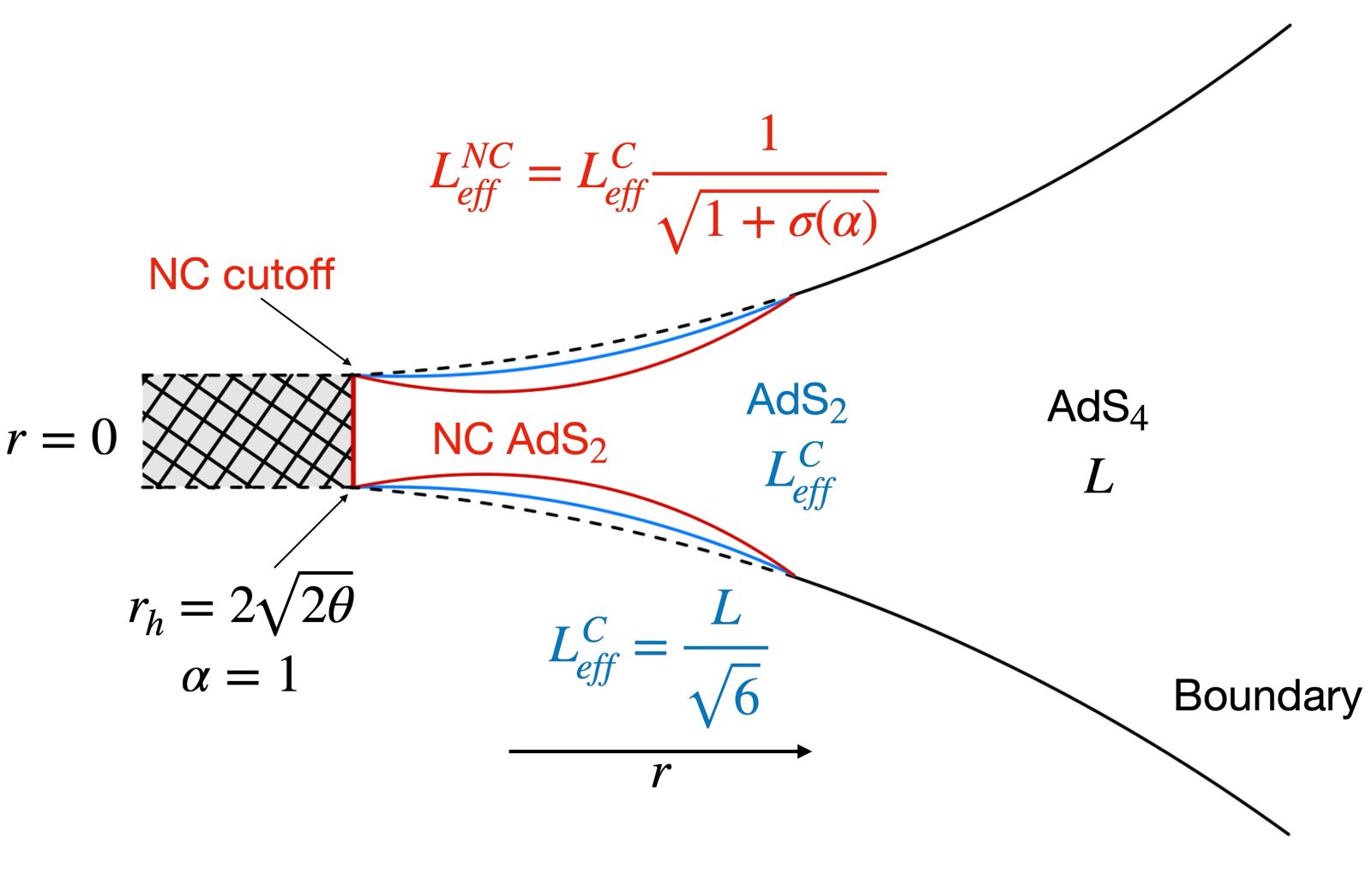}
    \caption{\begin{small}Schematic picture of the global structure of the $(3+1)$-dimensional dyonic NC solution (\ref{blackeningsol}). The effective NC curvature in the near-horizon geometry acquires contributions of the nearness parameter $\alpha$. The NC effects on the curvature start at the value $r_{h}=2\sqrt{2\theta}$ ($\alpha=1$) where the effective curvature for C ($L\effc=L/\sqrt{6}$) and for NC, coincide. If we increase the $\alpha$ parameter slightly greater than one, $L\effnc$ decreases (red curve) until it converges again to the effective curvature for the commutative AdS$_{2}$. Far away from the horizon radius, the bare $L$ for AdS$_{4}$ dominates (i.e., $\alpha\gg1$); however, there are still reminiscences from the NC mass distribution.
    \end{small}}
    \label{fig:globalgeom}
\end{figure}

Since the discretized confining box scenario for the study of holographic superconductors starts from $r_{h}$ (IR) to the AdS boundary (UV), the restriction $r_{h}^2\geq 8\theta$, is also satisfied for every point $r_{h}<r=\epsilon$ that defines the surface at which the dual dynamics takes place. However, there are memories of the NC effects even outside the horizon radius, a fact that has been used in other NC holographic constructions \cite{Pramanik:2015eka,Maceda:2019woa}. To reinforce these statements, in section \ref{section3}, we find the same nearness parameter when considering the NC effects on the Breitenlohner-Freedman bound for a perturbative scalar field and, a new consistent interpretation of the restriction $\alpha\geq 1$ relies on the effective curvature of the AdS$_2$ geometry: if $\alpha<1$ the effective curvature increases without bound, turning the geometry less curved, in conflict with the general knowledge about the near horizon geometry of the AdS charged black hole solutions \cite{Hartnoll:2008kx}. See Fig.~\ref{fig:globalgeom}.

\section{Thermodynamics of the bulk}\label{thermo}

In this section, we analyze the thermodynamic properties of the NC dyonic solution of the previous section and explore the NC effects on thermal quantities. Starting with the renormalized Euclidean on-shell action, the holographic dual to the free energy potential in the grand canonical ensemble
\begin{equation}
    S\ren=-\dfrac{1}{\kappa^{2}}\int d^{4}x\sqrt{-g}\left(\dfrac{6}{L^2}+\dfrac{L^2 \epsilon_{1}^{2}}{4}F^{2}\right)+\dfrac{1}{\kappa^{2}}\int d^{3}x\sqrt{-\gamma}\left( 2K+\dfrac{k}{L}\right),
\end{equation}
being $K=\gamma^{\mu\nu}K_{\mu\nu}$ the trace of the extrinsic curvature tensor $K_{\mu\nu}=-\dfrac{1}{2}\left( \nabla_{\mu}n_{\nu}+\nabla_{\nu}n_{\mu}\right)$ and $n_{\mu}$ an outward normal vector at some $r=\varepsilon$ slice with $\gamma$ the induced metric at that point. The constant $k$ is determined by the requirement to suppress the divergences when the surface defined by $r=\varepsilon$, tends to infinity\footnote{An interesting topological renormalization procedure called \emph{Kounterterms} is analyzed in \cite{Olea:2005gb,Olea:2006vd} for $d$ even and odd, respectively.}. The Gibbons-Hawking term is also present and is necessary to have a well-defined variational principle \cite{Balasubramanian:1999re,PhysRevD.15.2752}.

\begin{figure}[hbt!]
    \centering
    \includegraphics[width=0.5\textwidth]{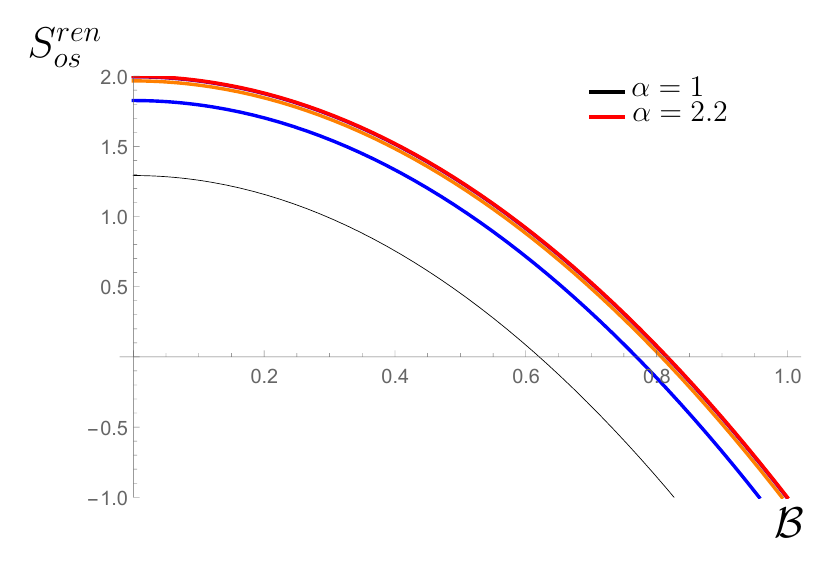}
    \caption{\begin{small} Noncommutative Euclidean renormalized on-shell action (taking $\mathcal{Q}=1$ and $r_{h}=1$) as a function of magnetic field $\mathcal{B}$. With $\alpha\gg1$, both actions coincide exactly, represented by the red curve. However, at $\alpha=1$, where the NC effects acquire more relevance, the on-shell action corresponds to the black curve. \end{small}}
    \label{fig:sosNC}
\end{figure}

With the above prescription, at the asymptotic AdS boundary, the renormalized Euclidean on-shell action turns out to be
\begin{equation}\label{sosren}
        S\ren = \dfrac{\beta_{T}\mathcal{V}_{2}}{\kappa^{2}}\dfrac{1}{L^{4}r_{h}}\left[ \left(2-\dfrac{\sqrt{\pi}}{2\tilde{\gamma}_{l}}\right)r_{h}^{4}+\left(2-\dfrac{\sqrt{\pi}}{2\tilde{\gamma}_{l}}\right)\mathcal{Q}^{2}-\left(2+\dfrac{\sqrt{\pi}}{2\tilde{\gamma}_{l}}\right)\mathcal{B}^{2}\right],
\end{equation}
where $\beta_{T}$ corresponds to the inverse of Hawking temperature and $\mathcal{V}_{2}$ is the $2$-dimensional volume in the transverse $(x_{1},x_{2})$ coordinates. The reader should note the appearance of the function
\begin{equation}
  \tilde{\gamma}_{l}\equiv\gamma_{l}\left(\dfrac{3}{2},2\alpha^{2}\right),
  \label{gammatilde}
\end{equation}
carrying the contributions of the on-shell action due to NC effects through $\alpha$ (\ref{thebound0}), defined in the previous section\footnote{Throughout the work, the lower gamma function evaluated in $2\alpha^{2}$ appears constantly hence, we adopt the convention $\tilde{\gamma}_{l}$ from now on.}. See Fig.~\ref{fig:sosNC}.

In the limit $\alpha\mapsto\infty$, the lower gamma function approximates to $\sqrt{\pi}/2$, therefore
\begin{equation}
    \lim_{\alpha\mapsto\infty}S\ren= \dfrac{\beta_{T}\mathcal{V}_{2}}{\kappa^{2}}\dfrac{1}{L^{4}r_{h}}\left(r_{h}^{4}+\mathcal{Q}^{2}-3\mathcal{B}^{2}\right),
\end{equation}
recovering the renormalized on-shell action of the commutative dyonic black hole, previously analyzed in \cite{refId0} with the same parametrization. The Eq. (\ref{sosren}) allows us to compute the thermodynamical variables, taking into account the thermodynamic potential $\mathcal{P}=-T\log\mathcal{Z}=-T S\ren$, namely\footnote{Recall that we identify the chemical potential with the horizon radius by the use of (\ref{qvsmu}).}
\begin{subequations}\label{termo}
   \begin{align}
    \rho &=-\dfrac{1}{\mathcal{V}_{2}}\dfrac{\partial \mathcal{P} }{\partial \mu}=\dfrac{1}{4}\dfrac{\epsilon_{1}^{2}}{\kappa^{2}}\left(4-\dfrac{\sqrt{\pi}}{\tilde{\gamma}_{l}}\right)\mu r_{h}=\dfrac{\epsilon_{1}}{2\kappa^{2}}\left(4-\dfrac{\sqrt{\pi}}{\tilde{\gamma}_{l}}\right)\dfrac{\mathcal{Q}}{L^{2}},\label{rho}\\ \nonumber \\
   \tilde{m} &=-\dfrac{1}{\mathcal{V}_{2}}\dfrac{\partial\mathcal{P}}{\partial\mathcal{B}}=-\dfrac{\epsilon_{1}}{2\kappa^{2}}\left(4+\dfrac{\sqrt{\pi}}{\tilde{\gamma}_{l}}\right)\dfrac{\mathcal{B}}{ r_{h}},\label{mag}\\ \nonumber \\
    s &= \dfrac{4\pi}{\kappa^{2}}\dfrac{r_{h}^{2}}{L^{2}},\label{entropy}  \\ \nonumber \\
    T &=\dfrac{3r_{h}^{4}-\mathcal{F}}{4\pi L^{2} r_{h}^{3}}-\mathcal{T}(r_{h},\mathcal{F};\alpha),\label{tempe}
  \end{align}    
\end{subequations}
corresponding to the charge density, magnetization, entropy density (calculated with Hawking-Bekenstein area law), and black hole temperature, respectively. In the last equation, we have defined
\begin{equation}\label{tempeNC}
    \mathcal{T}(r_{h},\mathcal{F};\alpha) \equiv \dfrac{\sqrt{2} }{\pi L^{2}}\dfrac{r_{h}^4+\mathcal{F}}{ r_{h}^3 }\dfrac{e^{-2 \alpha ^2} \alpha ^3}{ \tilde{\gamma_{l}}},
\end{equation}
such that its contribution to the Hawking temperature vanishes when $\alpha\gg 1$ ($r_{h}/2\sqrt{\theta}\gg 1$ ), reducing it to the dyonic commutative black hole\footnote{Even more, setting $\mathcal{F}=0$, the Hawking Temperature reduces to the ones of Schwarzschild black hole.}. Besides, the  energy-momentum tensor at the AdS boundary \cite{Balasubramanian:1999re}
\begin{equation}
    \langle T_{\mu\nu}\rangle = \dfrac{2}{\kappa^{2}}\left(K_{\mu\nu}-K \gamma_{\mu\nu}-\dfrac{2}{L}\gamma_{\mu\nu}\right),
\end{equation}
allows us to compute the energy density associated with a time translation symmetry
\begin{equation}\label{energyden}
    \dfrac{m}{\mathcal{V}_{2}}=\mathfrak{e} =\dfrac{1}{\kappa^{2}}\dfrac{\sqrt{\pi}}{\tilde{\gamma}_{l}}\dfrac{r_{h}^{4}+\mathcal{F}}{L^{4}r_{h}}.
\end{equation}
It is interesting to explore the local thermodynamic stability of the background (\ref{finalSol}) under the NC effects, codified by $\alpha$. To achieve this aim, consider the equation of state for the grand canonical ensemble (with $\mathcal{B}$ as an external parameter) $\mathfrak{e}=-\frac{1}{\mathcal{V}_{2}}TS\ren+Ts+\mu\rho$. Using the set of equations (\ref{rho})-(\ref{entropy}) on (\ref{energyden}) yields
\begin{equation}\label{eqofstate}
    \mathfrak{e} (\rho,s,\mathcal{B};\alpha) = \dfrac{\kappa}{16 L}\Upsilon_{1}(\alpha)\left(\dfrac{s}{\pi}\right)^{3/2}\left[1+\dfrac{16 \pi^{2}}{\kappa^{4}L^{2}}\left(\dfrac{\mathcal{B}}{s}\right)^{2}+16\pi^{2}\Upsilon_{2}(\alpha)\left(\dfrac{\rho}{s}\right)^{2}\right].
\end{equation}
The NC effects are arranged in the set $\lbrace\Upsilon_{i}(\alpha)\rbrace$, being
\begin{equation}
     \begin{aligned}
 \Upsilon _{1}\left ( \alpha\right )&\equiv \dfrac{e^{2\alpha^{2}}\left(\sqrt{\pi}+2\tilde{\gamma}_{l}\right)-8\sqrt{2}\alpha^{3}}{\tilde{\gamma}_{l}}e^{-2\alpha^{2}},\\
 \Upsilon_{2}\left ( \alpha  \right )&\equiv -\dfrac{\tilde{\gamma}_{l}^2 \left(e^{2\alpha^{2}}\left(3\sqrt{\pi}-10\tilde{\gamma}_{l}\right)+8 \sqrt{2} \alpha ^3\right)}{\left(\sqrt{\pi }-4 \tilde{\gamma}_{l}\right)^2 \left(e^{2\alpha^{2}}\left(\sqrt{\pi}+2\tilde{\gamma}_{l}\right)-8 \sqrt{2} \alpha ^3\right)},\\
\end{aligned}
\end{equation}
such that, in the limit $\alpha\mapsto\infty$ we have $\tilde{\gamma}_{l}\mapsto \sqrt{\pi}/2$ and the NC effects disappear progressively\footnote{There is a fast convergence to the commutative quantities when $\alpha\mapsto\infty$.} since
\begin{equation*}
    \Upsilon_{1}\mapsto 4, \qquad \Upsilon_{2}\mapsto \dfrac{1}{4},
\end{equation*}
and the exact equation of state of the commutative dyonic black hole is recovered \cite{refId0}.

\begin{figure}[h]
    \centering
    \includegraphics[width=0.5\textwidth]{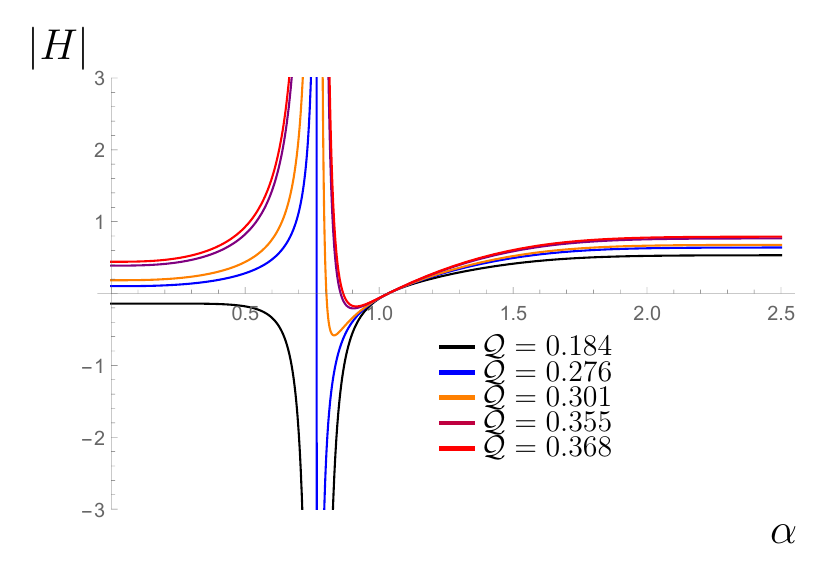}
    \caption{\begin{small} Determinant of the Hessian operator ($\vert H\vert$) as a function of $\alpha$ for different values of the electric charge. This function maintains its positivity above the value $\alpha\sim 1.05$. Note that this point acts as a \emph{fixed point}, no matter how the parameters are varied. \end{small}}
    \label{fig:hess}
\end{figure}

Regarding the possible instabilities, the condition over the Hessian: $\det\left[\partial_{(\rho,s)}^{2}\mathfrak{e} (s,\rho)\right]>0$, implies positive energies \cite{Hartnoll:2007ai}, i.e., a stable background. We go back to the black hole parameters $\left(\mathcal{Q},\mathcal{B},r_{h};\alpha\right)$ after computing the Hessian operator to get a better understanding of its behavior. Since the expression is quite formidable, in Fig.~\ref{fig:hess} we show the regions in which it maintains its positive-definite character. 

Positive values start around $\alpha=1$\footnote{A root of the curve $\vert H\vert$ vs $\alpha$.}, the point where the NC AdS confining box begins, as discussed in the previous section. The Hessian operator (determinant) is robust to changes in the values of the magnetic field but is sensitive to the horizon radius and the electric charge of the black hole. It is infinitely negative at $\alpha<1$ but can become finite (but still negative) if the electric charge and the horizon radius are comparable in magnitude. However, the latter case might violate the cosmic censorship condition. We stress that both cases occur for $\alpha<1$ and will not explore them further in this analysis. 

One would naturally have expected positive-definite values to start exactly at $\alpha=1$ and not at a slightly larger value, as indicated in Fig.~\ref{fig:hess}. We attribute this result to Hawking entropy (\ref{entropy}). Intuitively, the classical entropy obtained with the area law could acquire NC dependency such that, for $\alpha\gg1$, its corrections vanish. Accounting for these corrections to the Hawking entropy would require the use of the first law of Thermodynamics to relate the horizon radius and the black hole mass to the temperature. The resulting integration of $Tds=dM$ is not analytically possible in our setup. Numerical approximations have been developed in \cite{Banerjee:2008du}, for instance, showing indeed, a correction to the classical entropy at scales given by $\theta$.

Regarding the Maxwell field, we stated previously that the chemical potential $\mu$ (charge density $\rho$) gives the source (VEV) in the dual theory if we decide to work in the grand canonical (canonical) ensemble. Holography asserts that the electric component of the Maxwell field acts as a source field that fixes the charge density at the boundary \cite{zaanen_liu_sun_schalm_2015,Hartnoll:2016apf} 

\begin{equation}
    \rho\equiv\langle J^{t}\rangle=\dfrac{1}{\mathcal{V}_{2}\beta_{T}}\left(\dfrac{\delta S\os}{\delta A_{t}(r\mapsto\infty)}\right)=\dfrac{\epsilon_{1}^{2}}{4\kappa^{2}}\left(4-\dfrac{\sqrt{\pi}}{\tilde{\gamma}_{l}}\right)r_{h}\mu=\dfrac{\epsilon_{1}}{2\kappa^{2}}\left(4-\dfrac{\sqrt{\pi}}{\tilde{\gamma}_{l}}\right)\dfrac{\mathcal{Q}}{L^{2}},
\end{equation}
which is the same result (\ref{rho}) computed with the thermodynamic potential, representing a consistency check between pure gravity theory procedures and the holographic ones. To end this section, the first law of Thermodynamics
\begin{equation}
    d\mathfrak{e} =Tds-\mu d\rho,
\end{equation}
is satisfied using the set of the above-written formulae.
\subsection{Hawking temperature and extremality}
One of the main caveats that possess the general NC black hole solutions is that it is no longer possible to obtain in a closed form, an expression for the horizon radius and as a consequence, an exact zero temperature condition between electromagnetic charges, mass, and horizon radius, i.e., the so-called extremal black hole\footnote{More fundamentally, a censorship condition.}. In the case of the dyonic commutative black hole solution \cite{refId0}, obtained from the current construction when $\alpha\mapsto\infty$ ($\theta\mapsto 0$), the blackening factor reads
\begin{equation}\label{bknC}
    \chi\con(r)=1-\dfrac{\mathcal{M}}{r^3}+\dfrac{\mathcal{F}}{r^4}, \ \ \ \ \ \ \chi\extC (r)=1-\dfrac{4r_{0}^{3}}{r^{3}}+\dfrac{3r_{0}^{4}}{r^{4}},
\end{equation}
the extremal case $\chi\extC$ is reached when $3^3\mathcal{M}^4=4^4\mathcal{F}$ and $3r_{0}^4=\mathcal{F}$, being $r_{0}$ the merged horizon\footnote{Recall that $r_{0} $ corresponds to one single extremal horizon position, i.e., when $r_{-}=r_{+}=r_{0}$.}. It is well known that, in the extremal configuration, the background develops an AdS$_{2}\times\mathbb{R}^{2}$ structure near the horizon \cite{Gubser:2008px}
\begin{equation}
    ds^{2}\approx -\dfrac{6}{L^{2}}u^{2}dt^{2}+\dfrac{r_{0}^{2}}{L^{2}}d\Vec{x}^{2}+\dfrac{L^{2}}{6}\dfrac{du^{2}}{u^{2}},
\end{equation}
with $u=r-r_{0}$ and vanishing Hawking temperature $T\extC=0$. 

\begin{figure}[h]
    \centering
    \includegraphics[width=0.5\textwidth]{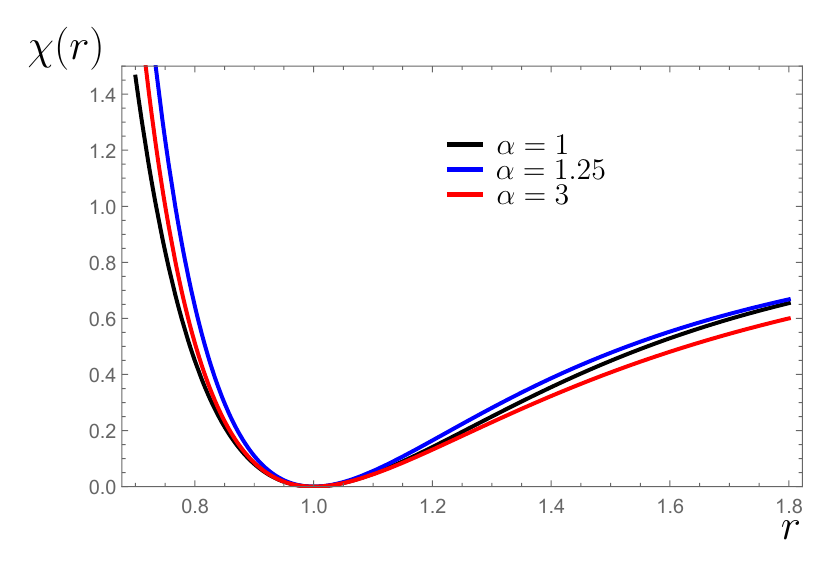}
    \caption{Fixing the horizon radius to unity, a comparison between extremal C (\ref{bknC}) and NC (\ref{bknNC}) blackening factors. The difference between the two functions seems to be minimal and, interestingly, for the blue curve, which corresponds to $\alpha\sim 1.25$, the difference between the functions is maximum, in contrast with the black curve for which $\alpha=1$. The convergence to the extremal C-function is very fast (red curve) at $\alpha=3$. The major difference occurs in the interval $r\in\left(1,1.8\right)$. Despite these minimal differences between C and NC functions, they have consequences in the holographic superconductor that we will explore in the next sections. }
    \label{fig:nhbkn}
\end{figure}

Regarding our NC solution, the fact that the Maxwell sources remain commutative gives consent to treat the extremal NC black hole in a closed form. First of all, recall the nearness parameter (\ref{thebound0}) to express the blackening factor as
\begin{equation}\label{bkn_a}
    \chi\Ncon (r) = 1+\dfrac{r_{h}^{3}}{r^{3}}\dfrac{\gamma\left(\tfrac{3}{2},\tfrac{2 \alpha^{2} r^2}{r_{h}^{2}}\right)}{\tilde{\gamma}_{l}}-\dfrac{\mathcal{F}}{r^{4}}\left(1-\dfrac{r_{h}}{r}\dfrac{\gamma\left(\tfrac{3}{2},\tfrac{2 \alpha^{2} r^2}{r_{h}^{2}}\right)}{\tilde{\gamma}_{l}}\right).
\end{equation}
Written in this form, $\alpha$ represents the NC deviations from the commutative solution due to a NC mass distribution that recovers the commutative character when $\alpha\mapsto\infty$. Calling for the Hawking temperature (\ref{tempe}) and equating it to zero (extremal case), allows us to solve for $\mathcal{F}$
\begin{equation}
    \mathcal{F}=r_{h}^{4}\left(3-\dfrac{32\alpha^{3}}{8\alpha^{3}+\sqrt{2}e^{2\alpha^{2}}\tilde{\gamma}_{l}}\right),
\end{equation}
and with it, rewrite (\ref{bkn_a}) as
\begin{equation}\label{bknNC}
    \chi\extNC (r)=1-\dfrac{4r_{h}^{3}}{r^{3}}\dfrac{\gamma\left(\frac{2 \alpha^{2} r^2}{r_{h}^{2}}\right)}{\tilde{\gamma}_{l}}\left(1-  \dfrac{8\alpha^{3}}{8\alpha^{3}+\sqrt{2}e^{2\alpha^{2}} \tilde{\gamma}_{l}} \right)+\dfrac{3 r_{h}^{4}}{r^{4}}\left(1-\dfrac{32\alpha^{3}}{3\left(8\alpha^{3}+\sqrt{2}e^{2\alpha^{2}} \tilde{\gamma}_{l}\right)}\right)
\end{equation}
which is the extremal blackening factor in the NC case. If $r$ is kept fixed, turns out that
\begin{equation}
    \lim_{\alpha\mapsto\infty} \dfrac{\gamma\left(\frac{2 \alpha^{2} r^2}{r_{h}^{2}}\right)}{\tilde{\gamma}_{l}}=1, \qquad \lim_{\alpha\mapsto\infty} \dfrac{\alpha^{3}}{8\alpha^{3}+\sqrt{2}e^{2\alpha^{2}} \tilde{\gamma}_{l}}=0,
\end{equation}
recovering the extremal case for the commutative black hole (\ref{bknC}).  See Fig.~\ref{fig:nhbkn}. 

The AdS$_{2}\times\mathbb{R}^{2}$ geometry solution of the commutative dyonic black hole represents the scenario for the dual deep IR quantum state of matter and was studied in \cite{Faulkner:2009wj, Faulkner:2010jy, Faulkner:2010gj} (and references therein) using a perturbative scalar field. In the next section, we shall explore the NC effects and their consequences regarding the minimal model of the holographic superconductor.

\section{AdS$_{2}$ and instabilities}\label{section3}
Over the constructed background (\ref{finalSol}), consider a perturbative massive scalar field 
\begin{equation}
    S_{\psi}=-L^{2}\epsilon_{2}^{2}\int d^{4}x\sqrt{-g}\left(\vert D\psi\vert^{2}+m^{2}\vert\psi\vert^{2}\right),
\end{equation}
that has been effectively decoupled from the total action (\ref{Action}) via $g_{s}\mapsto\infty$ in $\epsilon_{2}^{2}\equiv \kappa^{2}/g_{s}$\footnote{See subsection \ref{decoupling}. }. Nonetheless, the scalar field interacts with the commutative background electromagnetic charges via the minimal coupling: $D_{b}=\nabla_{b}-\textit{i}A_{b}$. If we assume a functional dependency of the form $\psi=\psi(x,y,r)$, the equation of motion acquires the form
\begin{equation}\label{scalarmastereq1}
    \partial_{r}\left[\dfrac{r^{2}}{L^{2}}U(r)\partial_{r}\psi\right]+\dfrac{r^{2}A_{t}^{2}}{L^{2}U(r)}\psi-\dfrac{m^2r^2}{L^2}\psi = 
    -\Delta_{(x,y)}\psi+2\textit{i}\left(\dfrac{2\mathcal{B}}{L^{4}}\right)x\partial_{y}\psi+\left(\dfrac{2\mathcal{B}}{L^{4}}\right)^{2} x^{2}\psi,
\end{equation}
being $\Delta_{(x,y)}$ the Laplacian operator in the transversal $(x,y)$-coordinates. It is licit to assume the functional dependency $\psi=e^{ipy}\gamma(x)R(r)$ under which the above equation is separable. Let $\lambda$ be the separation parameter, then (\ref{scalarmastereq1}) yields the following system of equations\footnote{Also we rewrite the electric component of the Maxwell field as $A_{t}=\frac{2\mathcal{Q}}{L^{2}r_{h}}\left(1-\frac{r_{h}}{r}\right)$, using (\ref{reparam1}). The same goes for the $B$-field.}
\begin{subequations}\label{separation}
	\begin{align}
        \partial_{r}\left(\dfrac{r^{2}}{L^{2}}U\partial_{r}R(r)\right)+\dfrac{1}{U}\left(\dfrac{r A_{t}(r)}{L}\right)^{2}R(r)-\left(\dfrac{mr}{L}\right)^{2}R(r) &=\left(\dfrac{2\mathcal{B}}{L^{4}}\right) \lambda R(r) \label{Req1}, \\ \nonumber\\
        -\left( \partial_{x}^{2}-p^{2}\right)\gamma (x) -2\left(\dfrac{2\mathcal{B}}{L^{4}}\right)px\gamma (x)+\left(\dfrac{2\mathcal{B}}{L^{4}}\right)^{2}x^2\gamma (x) &= \left(\dfrac{2\mathcal{B}}{L^{4}}\right) \lambda\gamma (x).\label{gammaeq1}
	\end{align}
\end{subequations}
The vortex structure is revealed by (\ref{gammaeq1}) since it is possible to bring it to a harmonic-oscillator-like equation, in a similar fashion as the Abrikosov procedure \cite{Abrikosov:1956sx}. Indeed, considering the change of coordinate $w=\sqrt{\mathfrak{b}}\left(x-p/\mathfrak{b}\right)$, with $\mathfrak{b}\equiv 2\mathcal{B}/L^{4}$, in this parametrization, the equation (\ref{gammaeq1}) can be recast in the form
\begin{equation}
    -\gamma''\left(w\right)+w^{2}\gamma\left(w\right)=\lambda\gamma\left(w\right),
\end{equation}
representing (a part of) the linearized equation that emanates from the Ginzburg-Landau free energy functional \cite{book:17888}. The solution of the above equation is generated by the Hermite polynomials
\begin{equation}
    \gamma_{n}\left(w;p\right)=e^{-w_{p}^{2}/4}H_{n}\left(w\right),
\end{equation}
with eigenvalues given by the discretized separation parameter $\lambda_{n}=2n+1$, being $n=0$ the \emph{Lowest Landau Level} (LLL). Abrikosov showed that, if we consider the product $C_{l}e^{ipy}\gamma_{n}\left(w\right)$ in the $\psi$ function, with $l$ an integer number, it is possible to construct a lattice by imposing periodicity in the ground state solution ($n=0$) codified by the normalization constant $C_{l}$, demonstrating by the way, that the free energy is minimal for the LLL with triangular shape of the fundamental cell in the lattice. We invite the reader to consult the holographic realization of the lattice structure, considering nonbackreacted matter in \cite{PhysRevD.106.L081902, Montull:2009fe, Maeda:2009vf, Albash:2009iq, Adams:2012pj, Srivastav:2023qof, Xia:2019eje} and with full backreaction in \cite{Donos:2020viz}.

Regarding the equation (\ref{Req1}) and, according to holographic renormalization, it gives us the scale on transversal $(w,y)$ directions (depending on the value of the holographic coordinate $r$) at which the dynamics take place. 

We are primarily interested in analyzing the above equations in the context of NC background. To achieve this aim, we stress the commutative backbones of AdS/CMT theory, which is based on the celebrated hair formation of the Reissner-Nordström black hole and related to spontaneous symmetry breaking interpreted as a phase transition from a vanishing scalar field to a non-trivial profile \cite{Gubser:2008px}. Moreover, it was shown in \cite{Hartnoll:2008kx} that the AdS$_{2}$ condensate is stable due to the strong gravitational attraction, even in the decoupled limit established in the lines above\footnote{What is more, with a neutral scalar field.}.  Guided by these works, we review the BF bound of the NC dyonic black hole described in the previous section. To expose the AdS$_{2}\times\mathbb{R}^{2}$ structure, we consider the expression for the blackening factor (\ref{bknG}). Expanding near the horizon, this function develops a double-zero, as shown in Fig.~\ref{fig:outerh}, due to the nonzero real roots (although is not possible to obtain these roots analytically)
\begin{equation}\label{extremalNCbh}
    ds^{2}\approx -(1+\sigma)\dfrac{6u^{2}}{L^{2}}dt^{2}+\dfrac{r_{h}^{2}}{L^{2}}d\Vec{x}^{2}+\dfrac{1}{(1+\sigma)}\dfrac{L^{2}}{6u^{2}}du^{2},
\end{equation}
where $u=r-r_{h}$. An AdS$_{2}\times\mathbb{R}^{2}$ topology arises with effective curvature radius
\begin{equation}\label{adsLeff}
    L\effnc\equiv l=\dfrac{L}{\sqrt{6}}\dfrac{1}{\sqrt{1+\sigma}},
\end{equation}
being
\begin{equation}\label{sigma}
    \sigma \equiv \dfrac{r_{h}^{3}\left(r_{h}^2-8\theta\right) }{6\left(4\theta ^{5/2}  \tilde{\gamma}_{l}\exp\left(r_{h}^{2}/4\theta\right)+ r_{h}^3 \theta\right) }, \ \ \ \ \ \ \tilde{\gamma}_{l}=\gamma_{l}\left(\frac{3}{2},\frac{r_{h}^{2}}{4\theta}\right),
\end{equation}
such that $\sigma$ vanishes when $\theta\mapsto 0$, recovering the AdS$_{2}\times\mathbb{R}^{2}$ geometry of the commutative dyonic black hole with effective curvature radius $L\effc=L/\sqrt{6}$. 

Over the extremal configuration (\ref{extremalNCbh}), the perturbative scalar field (\ref{Req1}) in near horizon geometry reduces to \cite{Horowitz:2009ij}
\begin{equation}
    R''(u)+\dfrac{2}{u}R'(u)-\dfrac{m\Ncon^{2}l^{2}}{u^{2}}R(u)=0,
\end{equation}
using the nearness parameter (\ref{thebound0}), the \emph{NC effective mass} turns out to be
\begin{equation}\label{massExtremal}
    m\Ncon^{2}=m^{2}-\left(1+\dfrac{32\alpha^{3}\left(\alpha^{2}-1\right)}{24\alpha^{3}+3\sqrt{2}e^{2\alpha^{2}}\tilde{\gamma_{l}}}\right)^{-1}\dfrac{2\mathcal{Q}^{2}}{3L^{2}r_{h}^{4}}+\dfrac{2\mathcal{B}\lambda}{L^{2}r_{h}^{2}},
\end{equation}
such that the expression for $m\Ncon^{2}$ reduces to the commutative one when $\alpha\mapsto\infty$ \cite{refId0}. The NC effective mass must obey
\begin{equation}\label{NCBF}
    \left(-\dfrac{9}{4}\right)_{\textit{AdS}_{4}}\leq m\Ncon^{2}l^{2}\leq \left(-\dfrac{1}{4}\right)_{\textit{AdS}_{2}},
\end{equation}
being the lower inequality the AdS$_{4}$ Breitenlohner-Freedman (BF) bound \cite{Breitenlohner:1982bm}. In the commutative case, it is well known that the value of the bare mass $m^{2}L^{2}=-2$ implies that the dominant modes in the UV boundary scalar field expansion are normalizable \cite{zaanen_liu_sun_schalm_2015,Witten:1998qj}.

\begin{figure}[ht]
		\centering
		\includegraphics[width=.5\textwidth]{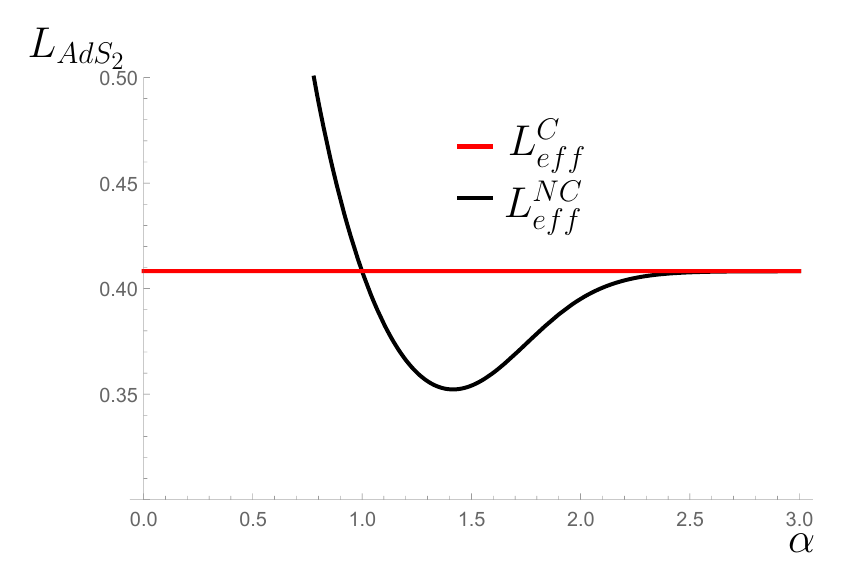}    
		\caption{The AdS$_{2}$ effective curvature radius for the NC ({black}) and C ({red}) cases as a function of $\alpha$; intersecting each other at the point $\alpha=1$. Below this value, $L\effnc$ increases without bound, representing a pathological geometry in which the near-horizon structure is less curved than the geometry far away from the horizon. In the opposite situation, when $\alpha\gg1$, both curvature radii converge (actually, converge quickly). There is a narrow interval in which the NC curvature is less than the C-case. Strong curvature will promote hair formation near the horizon relative to the C-case. In the next section, we shall explore this sentence and the effects on the holographic superconductor. }
    \label{fig: leff}
\end{figure}

At this point, we make contact with subsection \ref{alphasubsec} (and Fig.~\ref{fig:globalgeom}) regarding the nearness parameter. If $\alpha =1$, which is equivalent to take $r_{h}=2\sqrt{2\theta}$, the NC effective curvature radius (\ref{adsLeff}) reduces to the commutative case and the effective scalar field mass adopts the form
\begin{equation}\label{minimunM}
m\Ncon^{2}=m^{2}-\dfrac{2\mathcal{Q}^{2}}{3L^{2}r_{h}^{4}}+\dfrac{2\mathcal{B}\lambda}{L^{2}r_{h}^{2}}=m\con^{2}.
\end{equation}
Horizon radius less than $2\sqrt{2\theta}$ is forbidden due to the unbounded curvature radius below this value,  see Fig.~\ref{fig: leff}. Therefore, represents the minimum size that the NC background supports an AdS$_2$ structure in the near horizon geometry. Recall that, from the discussion of the Ricci scalar curvature of subsection \ref{alphasubsec}, $\alpha$ values less than one, would represent a horizon radius below the NC cutoff, and from the perspective of the effective AdS$_{2}\times\mathbb{R}^{2}$ geometry, implies an effective curvature radius greater than the bare one, $L$. Note also that, below the cutoff, the effective scalar field mass (\ref{massExtremal}) grows without bound, which is also a pathological behavior. In terms of the nearness parameter, the effective AdS$_{2}$ curvature can be cast in the form 
\begin{equation}
    L\effnc = \dfrac{L}{\sqrt{\dfrac{64 \left(\alpha ^2-1\right) \alpha ^3}{8 \alpha ^3+\sqrt{2} e^{2 \alpha ^2} \tilde{\gamma}_{l}}+6}},
\end{equation}
expression that will be useful in the next sections.
\section{Critical magnetic fields and the NC effects}\label{upperCmagnetic}
In this section, we study the behavior of the scalar field equation as a dual description of the type-II superconductor. First of all, we stress that our perturbative scalar field model over the dyonic NC background is unable to describe the thermal behavior (below a certain critical temperature) of the condensation parameter, dual to the density of superconducting state\footnote{In the next section, however, we analyze the existence of such a density of superconducting states.}.  Despite this limitation, it was argued in several works that this model certainly does make predictions regarding the behavior of the upper critical magnetic field  \cite{ Horowitz:2009ij,Albash:2008eh,refId0}. 

\begin{figure}[ht]
		\centering
		\includegraphics[width=.5\textwidth]{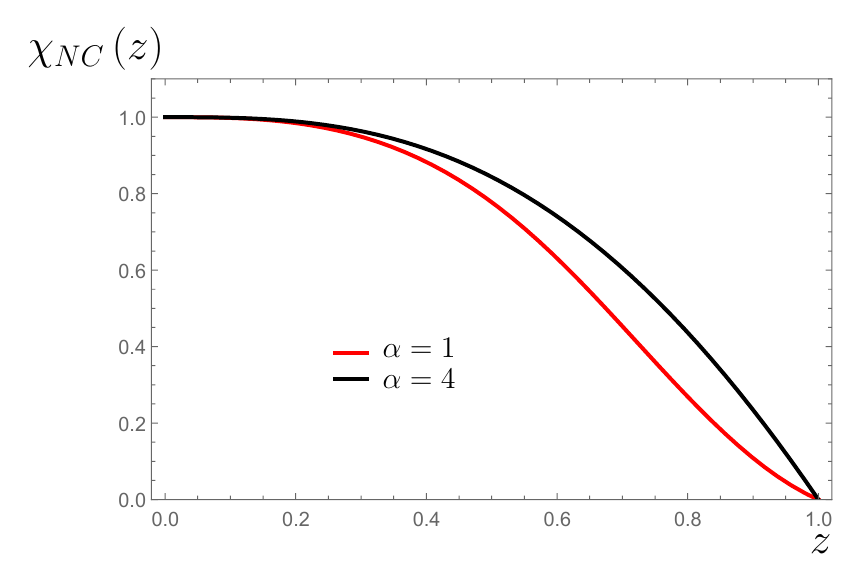}    
		\caption{Behavior of the two limiting cases of the blackening factor. As long as $\mathcal{F}$ remains small, both functions are positive definite in the $z\in\left[0,1\right]$ interval. }
    \label{fig: ap}
\end{figure}

In the language of the coordinate $z=r_{h}/r$, the radial part of the scalar field equation (\ref{Req1}), reads
\begin{align}\label{Rzeq}
   R''(z)+\left(\dfrac{\chi'\Ncon(z)}{\chi\Ncon(z)}-\dfrac{2}{z}\right)R'(z)&+\dfrac{L^2}{\chi\Ncon(z)}\left(\dfrac{L^2 A_{t}^{2}(z)}{r_{h}^{2}\chi\Ncon(z)}-\dfrac{m^{2}}{z^{2}}-\dfrac{2\mathcal{B}\lambda}{r_{h}^{2}L^{2}}\right)R(z) =0,\\ \nonumber \\
   A_{t}(z) &=\dfrac{2\mathcal{Q}}{L^{2}r_{h}}\left(1-z\right),
\end{align}
being $z=0$ and $z=1$ the location of the AdS boundary and the horizon, respectively. The blackening factor $\chi\Ncon(z)$ must be taken carefully since in the $r$-coordinate, the ratios $r_{h}^{2}/4\theta$ and $r^{2}/4\theta$ appear in the lower gamma functions and, under the $z$-coordinate we have 
\begin{align}\label{NCmetricz}
    \chi\Ncon(z) &=1-\dfrac{\gamma_{l}\left(\tfrac{3}{2},\tfrac{2\alpha^{2}}{z^{2}}\right)}{\tilde{\gamma}_{l}}z^{3}+\dfrac{\mathcal{F}}{r_{h}^{4}}\left(z-\dfrac{\gamma_{l}\left(\tfrac{3}{2},\tfrac{2\alpha^{2}}{z^{2}}\right)}{\tilde{\gamma}_{l}}\right)z^{3}.
\end{align}
Only by a limiting process, does the above function reach a finite value when $z\mapsto 0$. Also, we incorporate the nearness parameter (\ref{thebound0}) from now on in the blackening factor, representing the deviations due to NC effects of the commutative solution, i.e.,
\begin{equation}
    \lim_{\alpha\mapsto\infty}\chi\Ncon (z) = \chi\con (z).
\end{equation}
On the other hand, the radial equation (\ref{Rzeq}) for the normalizable modes associated with $m^{2}L^{2}=-2$, reduces to
\begin{equation}\label{Rzeq2}
    R''(z)+\left(\dfrac{\chi'\Ncon(z)}{\chi\Ncon(z)}-\dfrac{2}{z}\right)R'(z)+\dfrac{1}{\chi\Ncon(z)}\left(\dfrac{4\left(1-z\right)^{2}\mathcal{Q}^{2}}{r_{h}^{4}\chi\Ncon(z)}+\dfrac{2}{z^{2}}-\dfrac{2\lambda\mathcal{B}}{r_{h}^{2}}  \right)R(z)=0.
\end{equation}
The introduction of the above metric (\ref{NCmetricz}) allows us to explore two relevant situations, as it concerns the holographic superconductor. The first one corresponds to the limit $\alpha\gg 1$, describing a black hole with a horizon radius far away from the NC cutoff (\ref{thebound0}), in other words, horizon radius plenty greater than $\sqrt{\theta}$. The second one regards the opposite situation i.e., $\alpha\sim 1$, in which the black hole horizon radius is situated close to the NC cutoff, therefore with a very short length. These two limits are depicted in Fig.~\ref{fig: ap}. As a remark, the blackening factor for both of these limiting cases, truly codifies the entire spacetime in the $z\in [0,1]$-interval, as long as $\mathcal{F}$ remains small.  The case $\alpha\sim 1$ will be the one that allows us to explore in the next section, the IR regime of the holographic superconductor and we expect the NC corrections to be relevant there.

At the UV boundary, the solution of the $R$-equation (\ref{Rzeq2}) behaves as
\begin{equation}\label{bdyExp}
    R(z\mapsto 0 )= J_{+}z^{\Delta_{+}}+J_{-}z^{\Delta_{-}},
\end{equation}
where $\Delta_{\pm}=\frac{1}{2}\left(3\pm\sqrt{9+4 L^{2}m^{2}}\right)$ ($\Delta_{+}=2,~\Delta_{-}=1$ for $m^{2}L^{2}=-2$)  are the conformal dimensions of the dual operator $\mathcal{O}$ to the scalar field $\psi$, being $\Delta_{-}$ the leading (non-normalizable) order associated to the source and, $\Delta_{+}$ the subleading (normalizable) order associated to the vacuum expectation value (VEV). Performing a series expansion at the horizon $z=1$, the solution of (\ref{Rzeq2}) has the general form
\begin{equation}\label{nhExp}
    R(z)=R(1)+R'(1)(z-1)+\dfrac{1}{2}R''(1)\left(z-1\right)^{2}+...\ .
\end{equation}
The matching method allows us to obtain an approximate solution of (\ref{Rzeq2}) by plugging, at some $z=z_{m}\in [0,1]$ the asymptotic solutions (\ref{bdyExp}) and (\ref{nhExp}). The difficulties arise in obtaining the coefficients of the above expansion due to NC contributions. To deal with this technical issue, it is convenient to keep the expansions in a general form, neglecting the logarithmic divergence at $z=1$
\begin{align}\label{coefnh}
    R'(1) &=\dfrac{1}{\chi\Ncon '\left ( 1 \right )}\left [ m^{2}L^{2}+\frac{2\mathcal{B}\lambda }{r_{h}^{2}} \right ]R(1),\\ 
    R''(1) &=-\frac{1}{2\chi\Ncon '\left ( 1 \right )}\left [ \left (\chi\Ncon ''\left ( 1 \right )-2\chi\Ncon '\left ( 1 \right ) -m^{2}L^{2}-\frac{2\mathcal{B}\lambda}{r_{h}^{2}} \right )R'\left ( 1 \right )+\left ( 2m^{2}L^{2}+\frac{L^{4}A_{t}'^{2}\left ( 1 \right ) }{r_{h}^{2}\chi\Ncon '\left ( 1 \right )} \right )R\left ( 1 \right ) \right ]\nonumber.
\end{align}
With the above coefficients, the expansion (\ref{nhExp}) (up to second order) matched with (\ref{bdyExp}) at $z=z_{m}$, allows us to construct a \emph{phase space} parameterized by the set $\lbrace \mathcal{Q},\mathcal{B},\alpha ; z_{m}\rbrace$\footnote{Recall that, as we mentioned in Sec. \ref{section3}, the parameter $z_{m}$ (coordinate $r$ before the diffeo $z=r_{h}/r$) has the interpretation of the renormalization scale from UV at the boundary to IR close to the horizon \cite{zaanen_liu_sun_schalm_2015}.}. It is well known that the value $m^2 L^2=-2$ has normalizable falloffs near the boundary therefore, both modes can be interpreted as the dual operator of the superconducting condensation order parameter because both asymptotic behaviors satisfy the BF bound\footnote{In the language of holography, both normalizable modes are reflected in the freedom to choose the \emph{quantization scheme} \cite{Witten:1998qj}. } (\ref{NCBF}). In this work, we deal with the mode $\Delta_{+}=2$, setting the source $J_{-}=0$, as the spontaneous symmetry-breaking demands. 

From what is described in the above lines, and the use of the expansions (\ref{bdyExp}) and (\ref{nhExp}), we can write down the expressions that capture the phase space of the superconductor, having the structure
\begin{equation}\label{phaseS}
\begin{array}{rcl}
  0   &=& r_{h}^{2}a_{3}+a_{2}\mathcal{B}+\dfrac{1}{r_{h}^{2}}\left(a_{1}\mathcal{B}^{2}-a_{4}\mathcal{Q}^{2}\right)-r_{h}^{2}\dfrac{J_{+} z_{m}^{2}}{R(1)}, \\  \\
  0   &=&  r_{h}^{2}b_{3}+b_{2}\mathcal{B}+\dfrac{1}{r_{h}^{2}}\left(b_{1}\mathcal{B}^{2}-b_{4}\mathcal{Q}^{2}\right)-2r_{h}^{2}\dfrac{J_{+}z_{m}}{R(1)},
\end{array}
\end{equation}
being the set $\lbrace a_{i}=a_{i}(\alpha,z_{m}),b_{i}=b_{i}(\alpha,z_{m})\rbrace$ complicated functions of $\alpha$ and the point match $z_{m}$, detailed in the appendix \ref{coefficients}.

The horizon radius, related to the Hawking temperature (\ref{tempe}), depends on $\alpha$ and measures the deviations in the commutative temperature due to the proximity to the NC cutoff. For small electromagnetic charges \cite{Hartnoll:2016apf}, we consider
\begin{equation}\label{Hawn}
  T\h = \dfrac{3r_{h}}{4\pi L^{2}}\left(1-\dfrac{\sqrt{2}\alpha^{3}e^{-2\alpha^{2}}}{2\tilde{\gamma}_{l}}\right)\equiv T_{0}-T_{\alpha},
\end{equation} 
such that, when $\alpha\mapsto\infty$ the Hawking temperature reduces to the one associated with a neutral black hole $T_{0}$ ($T_{\alpha}\mapsto 0$), corresponding to the high-temperature superconductor since the horizon radius is much greater than $\sqrt{\theta}$. The phase space associated with the full temperature $T\h$ considers the NC regime. The former is extensively studied in several works, see, for instance, \cite{Hartnoll:2007ip} in the context of commutative Abelian-Higgs holographic superconductivity, \cite{Pramanik:2015eka,Maceda:2019woa} for the NC versions and \cite{Albash:2008eh,refId0} using the perturbative scalar field.

Using the expressions (\ref{phaseS}), we solve for the magnetic field as a function of $(\mathcal{Q},\alpha,z_{m},r_{h})$ followed by the use of (\ref{Hawn}) to express the horizon radius as a function of the temperature. This procedure allows us to write down a formidable expression for the magnetic field in the canonical ensemble, namely\footnote{But otherwise exact, up to the matching method concerns.}
  \begin{align}\label{magneticF}
    \mathcal{B} &=\dfrac{1}{(2a_{1}-b_{1}z_{m})\left(4\sqrt{2}\alpha^{3}-3e^{2\alpha^{2}}\tilde{\gamma}_{l}\right)^{2}} \left[ \vphantom{\left.+ (2 a_1-b_1 z_{m})(2  a_4 -b_4 z_{m})\mathcal{Q}^2\left(4 \sqrt{2} \alpha ^3-3 e^{2 \alpha ^2} \tilde{\gamma}_{l}\right)^4     \right]^{1/2}}-8e^{4\alpha^{2}}\left(2a_{2}-b_{2}z_{m}\right)\tilde{\gamma}_{l}^{2}L^{2}\pi^{2}T^{2}\right.\nonumber \\ 
    &\quad+ \left( \vphantom{+ (2 a_1-b_1 z_{m})(2  a_4 -b_4 z_{m})\mathcal{Q}^2\left(4 \sqrt{2} \alpha ^3-3 e^{2 \alpha ^2} \tilde{\gamma}_{l}\right)^4} 64 \pi ^4 e^{8 \alpha ^2} L^8 T^4 \tilde{\gamma}_{l}^4 \left((b_2 z_{m}-2 a_2)^2-4 (2 a_1-b_1 z_{m})(2a_3-b_3 z_{m})\right)\right.\nonumber\\
    &\quad \left.\left.+ (2 a_1-b_1 z_{m})(2  a_4 -b_4 z_{m})\mathcal{Q}^2\left(4 \sqrt{2} \alpha ^3-3 e^{2 \alpha ^2} \tilde{\gamma}_{l}\right)^4     \right)^{1/2} \right].
\end{align}

We stress that our construction allows us to treat the dual system as a Type-II superconductor due to the London gauge in the solution of Einstein equations implying the existence of harmonic oscillator-like equation (\ref{gammaeq1}). Therefore, (\ref{magneticF}) defines the \emph{upper critical magnetic field} $\mathcal{B}_{c2}$, using the fact that the superconducting state and the magnetic field vanish at some temperature
\begin{align}\label{TcritCE}
    T_{c}(\mathcal{Q},\alpha;z_{m}) =\dfrac{e^{- \alpha ^2}\sqrt{\tilde{\gamma}_{l}\left(8\sqrt{2}\alpha ^{3}-6e^{2 \alpha ^2} \tilde{\gamma}_{l}\right)}\quad \vert z_{m}-1\vert^{1/4}}{\left[ e^{2 \alpha ^2}L^2m^2\tilde{\gamma}_{l}\left( \frac{(2 z_m-3)}{\left(6e^{2 \alpha ^2} \tilde{\gamma}_{l}-8\sqrt{2}\alpha ^{3}\right)}+\frac{(z_m-1)\left(e^{2 \alpha ^2}L^2m^2 \tilde{\gamma}_{l}-16\sqrt{2}\alpha ^{5}\right)}{\left(6e^{2 \alpha ^2} \tilde{\gamma}_{l}-8\sqrt{2}\alpha ^{3}\right)^2} \right)-1\right]^{1/4}} \dfrac{\sqrt{\mathcal{\vert Q\vert}}}{2\sqrt{2}\pi L^{4}}.
\end{align}

\begin{figure}[h]
    \centering
    \begin{subfigure}[b]{.49\textwidth}
    \centering
    \includegraphics[width=0.95\textwidth]{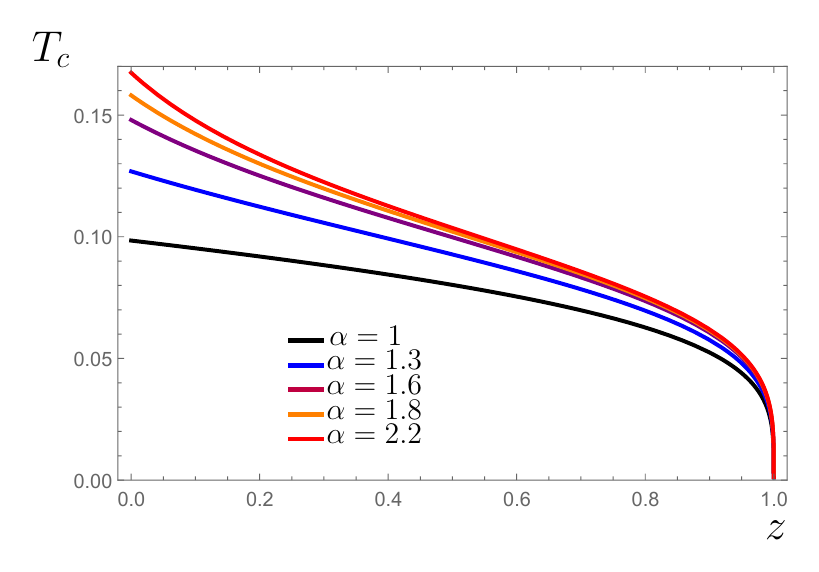}
    \end{subfigure}
    \hfill
    \begin{subfigure}[b]{.49\textwidth}
		\centering
		\includegraphics[width=.95\textwidth]{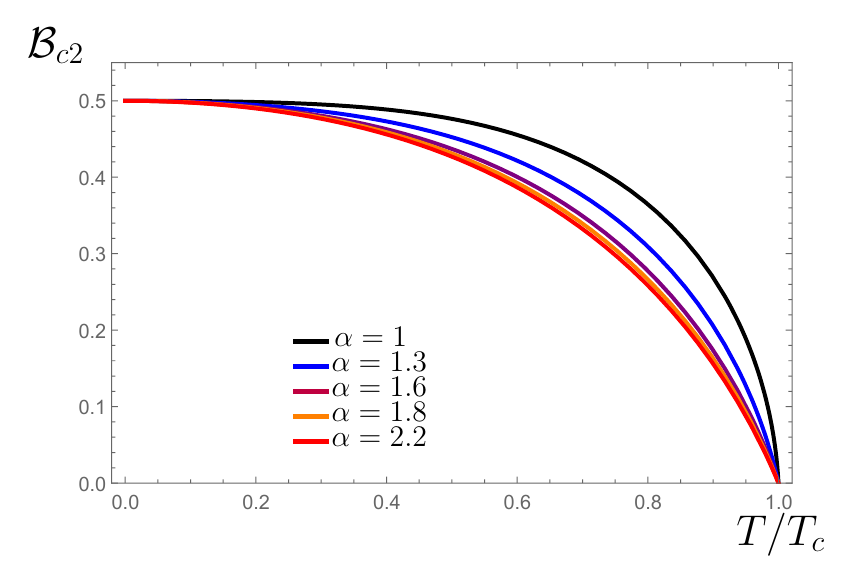}    
	\end{subfigure}
    \caption{\begin{small}Both figures are referred to the canonical ensemble for $\langle\mathcal{O}_{2}\rangle$ in which the charge density is fixed at the boundary, taken $\mathcal{Q}=\frac{1}{2}$ and $L=1$. \emph{Left}: Critical temperature (\ref{TcritCE}) as a function of the point match $z_{m}$ and different values of $\alpha$. The NC effects ($\alpha\sim 1$) decrease the critical temperature at the UV ($z_m\sim 0$) while disappearing progressively in the IR ($z_{m}\mapsto 1$). \emph{Right}: Normalized thermal behavior of the critical magnetic field at $z_{m}=0.522$. In both figures, the NC effects quickly disappear when $\alpha$ increases, and converge exactly at the commutative case analyzed in \cite{refId0}. We invite the reader to confront Fig. 7 of \cite{Pramanik:2015eka} with the present figure. \end{small}}
     \label{fig: Censemble}
\end{figure}

With the above formula, (\ref{magneticF}) can be normalized. Unfortunately, the resulting expression is quite formidable and we were unable to write it down in this report. See Fig.~\ref{fig: Censemble}.

\begin{figure}[ht]
		\centering
		\includegraphics[width=.5\textwidth]{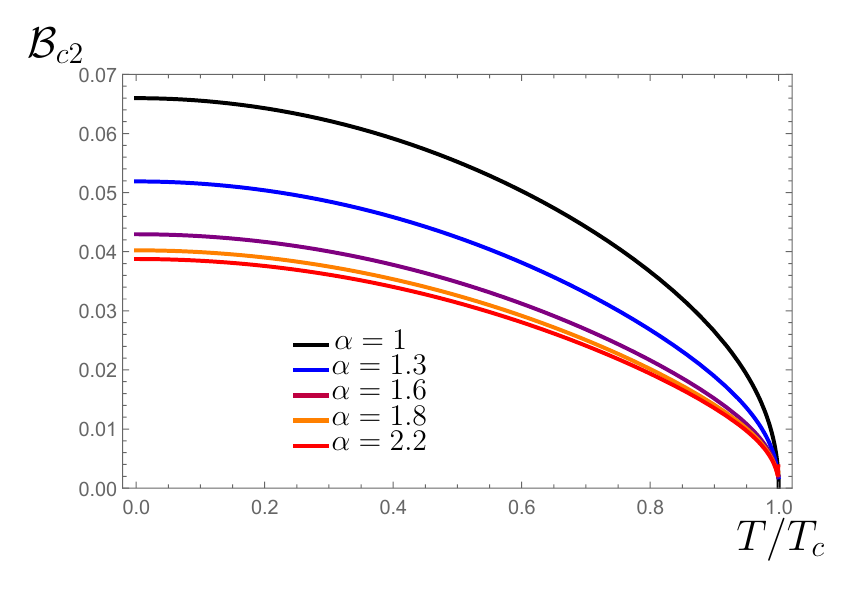}    
		\caption{\begin{small}The shape of the upper critical magnetic field in the grand canonical ensemble close to the horizon. As $\alpha$ increases, the amplitude of the field also increases; therefore, NC effects act in favor of the amplitude of $\mathcal{B}_{c2}$ of the type-II superconductor. We take $z_{m}=0.999$ and $L=1$, $\mu=\frac{1}{2}$. Again, as $\alpha$ increases, the curve exactly converges to the C-case \cite{refId0}.\end{small} }
    \label{fig: GCensemble}
\end{figure}

Following the same arguments, we can construct the equivalent phase space expressions (\ref{phaseS}) for the grand canonical ensemble in which the chemical potential $\mu=2\mathcal{Q}/L^2 r_{h}$ is fixed at the boundary. This case adds more complexity due to an extra contribution to the Hawking temperature, considerably raising the order of the analogous expressions (\ref{phaseS}) hence, a further approximation needs to be taken into account. The strategy is to restrict the blackening factor (\ref{NCmetricz}) close to the horizon $z=1$ up to first order, which allows us to absorb the NC effects, codified in the lower gamma functions. Once we obtain the phase space structure, we return the absorbed NC parameter to the phase space. Although the shape obtained in this case is restricted to $z_{m}\sim1$, we had been able to obtain the thermal behavior of the critical magnetic field that converges exactly to the C-case\footnote{In \cite{refId0}, the gran canonical shape of the normalized thermal behavior of critical magnetic field for every $z_{m}$, was successfully obtained. We invite the reader to consult this previous work.}. Within these considerations, in Fig.~\ref{fig: GCensemble}, we show the thermal behavior for the critical magnetic field and different values of $\alpha$. Again, fast convergence to the C-case is observed as $\alpha$ increases.

The reader should note in Fig.~\ref{fig: GCensemble} that, an interesting phenomenon occurs in the grand canonical ensemble: the critical magnetic field grows its amplitude as we increase the NC effects ($\alpha\sim 1$) for fixed values of the chemical potential. For the canonical ensemble case (Fig.~\ref{fig: Censemble}), we also observe an enhanced profile of the upper critical magnetic field but the amplitude remains the same, as $\alpha$ is varied. In both cases, the shapes reduce to the C-case as we increase the nearness parameter.

Another interesting result is that we were able to write down an amenable exact expression for the critical temperature in the grand canonical ensemble
\begin{equation}
T_{c}= 
\dfrac{ e^{-2 \alpha ^2} \sqrt{(3e^{2 \alpha ^2}  \tilde{\gamma}_{l}-4 \sqrt{2} \alpha ^3)(e^{2 \alpha ^2}\tilde{\gamma}_{l}+4 \sqrt{2} \alpha ^3)}  }{  \tilde{\gamma}_{l}  }\dfrac{\mu}{8\pi},
\end{equation}

that reduces to $\sqrt{3}\mu/8\pi$ when $\alpha\gg1$, reported previously in \cite{refId0}.


Even though the NC effects seem to be very small (but otherwise considerable), it is worth mentioning that our construction is completely analytical, up to the matching method concern, through a continuously varying NC parameter. As far as the authors know, these kinds of typical curves in holographic superconductivity with analytical NC effects, have not been reported previously.

To end this section, the above results show the gravitational description of the critical magnetic field. Two more quantities embrace the holographic description, namely, the conductivity and the thermal behavior of the condensate. As we stated in the introduction, the latter quantities can not be captured with our model since the Maxwell fields do not interact dynamically with the scalar field. In the next section, we analyze a Schrödinger equation for the scalar field that gives us robust evidence for the existence of a condensed dual-order parameter. 
\section{Schrödinger potential in the NC near horizon}\label{Spotential1}
In this section, we address the NC effects on the near-horizon geometry taking into account the nearness parameter (\ref{thebound0}) and their consequences on the condensation of the scalar field. As discussed in Section \ref{section3}, the near-horizon geometry of the NC dyonic black hole exhibits an AdS$_2\times\mathbb{R}^{2}$ topology. By virtue of the effective curvature AdS radius (which is stronger than the commutative configuration), we expect that a scalar field condenses in the form of hair due to gravitational pulling inward. We give an alternative analysis (relative to the Higgs-Abelian holographic theory \cite{Hartnoll:2008vx}) in terms of the bound states generated by the Schrödinger potential associated with the scalar field, which proves the existence of hair and justifies the profile of the critical magnetic fields of the previous section, typical in the Type-II superconducting systems. Recall that the Schrödinger potential captures the bound states of a given system, and if the spectrum is quantized, there is a relation between the depth of the potential and the number of bound states that it contains. Therefore, if the NC effects produce deeper potentials relative to the commutative case in the near-horizon geometry, we need to exhibit that there are more eigenstates captured. To do so, a Schrödinger equation with well-defined eigenvalues must be constructed. This procedure is what we are to perform next.

Taking into account that the radial equation (\ref{Rzeq}) codifies the renormalization group of the dual system, the transformation $R(z)\mapsto s(z)\mathcal{Z}(z)$ with the rescaling
\begin{equation}\label{rescale}
    s(z) = \dfrac{z}{\left(\chi\Ncon (z)\right)^{1/4}},
\end{equation}
transforms the $R$-equation into 
\begin{equation}\label{protoSch}
    -\sqrt{\chi\Ncon} \dfrac{d}{d z}\left( \sqrt{\chi\Ncon} \mathcal{Z}'\left(z\right) \right)+V\left(z\right)\mathcal{Z}\left(z\right)=E_{n}\mathcal{Z}\left(z\right).
\end{equation}
If we define a coordinate such that $\partial_{z^{*}}=\sqrt{\chi\Ncon}\partial_{z}$, the above expression can be recognized as a time-independent Schrödinger equation with potential\footnote{See \cite{zaanen_liu_sun_schalm_2015, Hartnoll:2016apf} for the construction of Schrödinger potential in case of time-dependent scalar field and \cite{Iqbal:2011in, Hartnoll:2011dm, Cubrovic:2011xm} for Dirac field towards the question of the holographic Fermi surface in the quantum criticality regime, and the holographic Fermi groundstate.} 
\begin{equation}\label{uniqueS}
    V\Ncon\left(z^{*}\right) = -\dfrac{4 \mathcal{Q}^2 (z^{*}-1)^2}{r_{h}^4 \chi\Ncon(z^{*})}-\dfrac{\chi\Ncon'(z^{*})}{z^{*}}-\dfrac{\chi\Ncon'(z^{*})^2}{16 \chi\Ncon(z^{*})}+\dfrac{\chi\Ncon''(z^{*})}{4}+\dfrac{2 \chi\Ncon(z^{*})}{z^{*^2}}+\dfrac{L^2 m^2}{z^{*^2}},
\end{equation}
and energy spectrum
\begin{equation}\label{eigenenergy}
    E_{n}=-\dfrac{2\mathcal{B}\lambda_{n}}{r_{h}^{2}}.
\end{equation}
The discretized energy values are related to the Hermite polynomials, which span the solution space of the equation (\ref{gammaeq1}) being $E_{0}$ the LLL (see section \ref{section3}). 

The perturbative scalar field with energy $E_{n}$ experiences the NC confining box by virtue of the curvature and the electromagnetic charges of the bulk. Furthermore, its dynamics are also strongly dependent on its mass. As we stated in the introduction of this section, we look for bound states of the potential as evidence of the existence of Cooper pairs in the dual superconductor\footnote{The dual density of Cooper pairs is the norm of the complete scalar field $\vert \psi\vert^{2}$.}. If $z^{*}_c$ corresponds to the local minimum of the potential well then, the neighborhood around it and the turning points for a given energy, define the region where bound states exist. Although the potential (\ref{uniqueS}) is valid for all values of $(\mathcal{B},\mathcal{Q},r_{h})$, there is a strong restriction that comes from the extremality condition. In our NC configuration, it is not possible to obtain in a closed form the analytical expression for extremal condition however, we find it convenient to take valid the extremal condition for the commutative configuration (that we called \emph{C-extremal} from now on), explored in \cite{refId0}. A fine-tuning in the plots can be achieved if we parametrize the electromagnetic charges as $\mathcal{B}=\sqrt{\nu}\sin\kappa$ and $\mathcal{Q}=\sqrt{\nu}\cos\kappa$. In this parametrization, $\nu\in (0,3)$ represents an \emph{extremality parameter} being $\nu=3$ the C-extremal case and $\nu=0$ the neutral black hole. The $\kappa$ parameter, on the other hand, controls the relative values between electromagnetic charges, e.g., $\kappa=0$ corresponds to the Reissner-Nordström black hole, without a magnetic charge hence, with energy $E_{n}=0$ for all $n$.

Since the coordinate transformation $z \mapsto z^{*}$ can not be integrated analytically
\begin{equation}\label{zstarc}
    z^{*}=\int^{z}\dfrac{dz'}{\sqrt{\chi\Ncon\left(z'\right)}},
\end{equation}
we perform an approximate expression near the horizon $z=1$, yielding
\begin{equation}\label{zstar}
    z^* \approx -\frac{2r_{h} ^{2}e^{\alpha ^{2 }}\tilde{\gamma }_{l}^{^{\frac{1}{2}}}}{\sqrt{\left ( 3 r_{h}^{4}-\nu \right )\tilde{\gamma }_{l}-4\sqrt{2}\alpha ^{3}\left ( r_{h}^{2}+\nu  \right )}}\sqrt{1-z}, \qquad \alpha^{3}< \dfrac{\left(3 r_{h}^{4}-\nu\right)}{4\sqrt{2}\left(r_{h}^{2}+\nu\right)}\tilde{\gamma}_{l}.
\end{equation}
In terms of the star coordinate, the horizon is located at $z^{*}=0$. We expect that the NC effects become relevant at the extremality and when $m^{2}L^{2}$ are close to the upper BF bound (\ref{NCBF}) in the sense that the stepper potential well is capable of capturing more bound states. Substituting (\ref{zstar}) into (\ref{uniqueS}), all the physics is codified in the Schrödinger potential. Unfortunately, its final expression is quite involved and we do not write it down in this report. Nonetheless, the structure of the potential (\ref{uniqueS}) in the near-horizon and extremality conditions is analyzed as a function of the nearness parameter, as we shall see now.

\begin{figure}[htp]
	\centering
	\begin{subfigure}[b]{.496\textwidth}
		\centering
		\includegraphics[width=.95\textwidth]{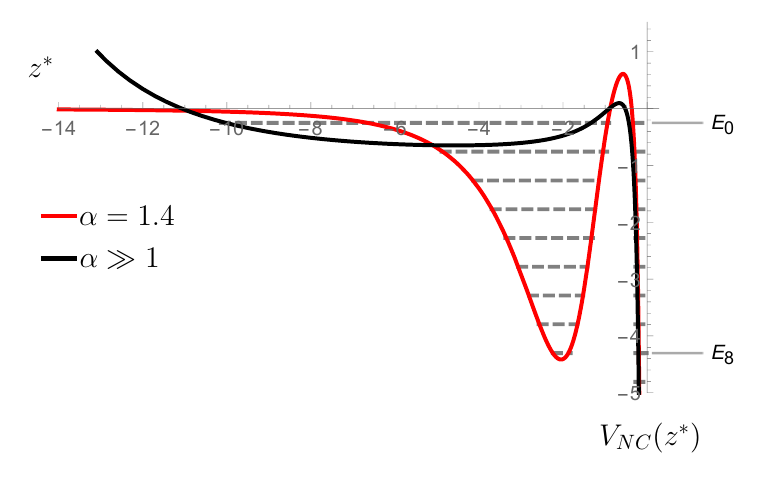}
		\caption{}
		\label{fig:newPot1a}
	\end{subfigure}
	\hfill     
	\begin{subfigure}[b]{.496\textwidth}
		\centering
		\includegraphics[width=.95\textwidth]{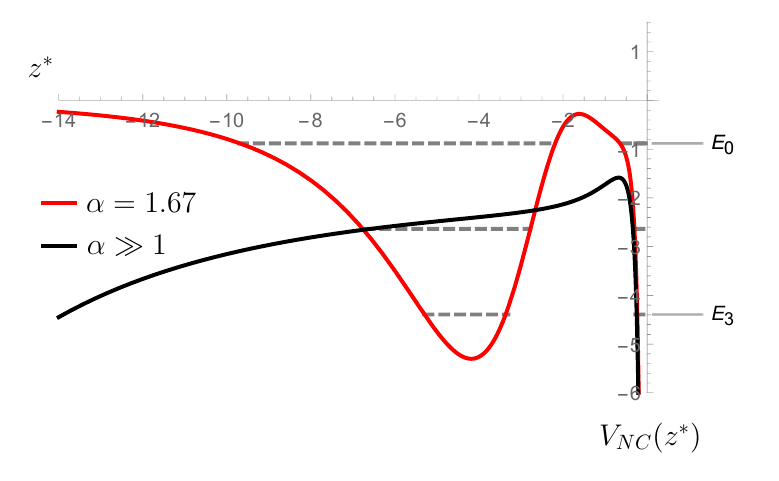}    
		\caption{}  
		\label{fig:newPot1b}
	\end{subfigure}
	\caption{\begin{small}NC Schrödinger potential (\ref{uniqueS}) for different values of the nearness parameter $\alpha$. Both cases are related to the near extremality $\nu=2.99$ dyonic black hole. (a) The near extremality NC potential for a mass close to AdS$_{2}$ BF bound, $m^2 L^2=-0.276$. We take $\kappa =0.073$ rad i.e., the electric charge has more strength than the magnetic one. (b) The near extremality NC potential for the normalizable mode $m^{2}L^{2}=-2$. We take $\kappa =0.26$ rad. As it can be seen, the NC effects promote the stability of this mode in the IR dual superconductor. \end{small}}
    \label{fig:newpot1}
\end{figure}

For the sake of clarity on the relation between the depth of the formed Schrödinger potential well and the number of bound states that captures, two eigenvalues of energy $E_{n}$  of the expression (\ref{eigenenergy}), starting with the LLL, are marked in Fig. \ref{fig:newpot1}.  We consider first the system in the near extremal case and small magnetic field, as well as a scalar field mass close to the BF bound for AdS$_{2}$, a situation depicted in Fig. \ref{fig:newPot1a}. The NC effects in the near horizon and near extremal geometry are controlled by the nearness parameter $\alpha$ and increase the number of bound states reflected in the more pronounced potential wells close to the cutoff $\alpha =1$. In other words, the stepper NC effective curvature in the AdS$_{2}$ near-horizon region, promotes the proliferation of bound states dual to the density of Cooper pairs, demonstrating that the NC background acts in favor of hair formation relative to the C-extremal background. Interestingly, the case of the normalizable mode $m^{2}L^{2}=-2$ of the figure \ref{fig:newPot1b}, shows that the commutative case $\alpha\gg 1$ can not form a bound state and no hair formation is present. However, turning on the NC effects $\alpha\mapsto 1$, robust potential wells start to form, retaining bound states, including the LLL. Both cases constitute a novel result since, it was pointed out in previous literature that this particular mode is unstable in the IR, nonetheless stable at the UV \cite{Faulkner:2010jy, Iqbal:2010eh}.

\begin{figure}[h]
	\centering
	\begin{subfigure}[b]{.496\textwidth}
		\centering
		\includegraphics[width=.95\textwidth]{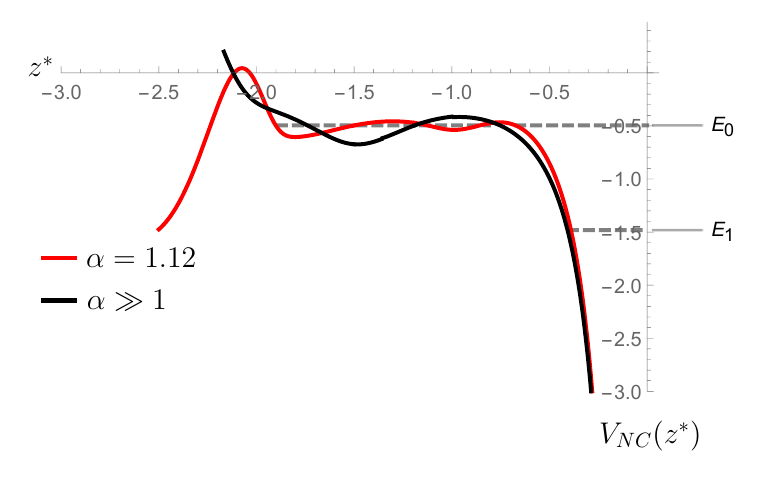}
  \caption{}
		\label{fig:newPot2a}
	\end{subfigure}
	\hfill     
	\begin{subfigure}[b]{.496\textwidth}
		\centering
		\includegraphics[width=.95\textwidth]{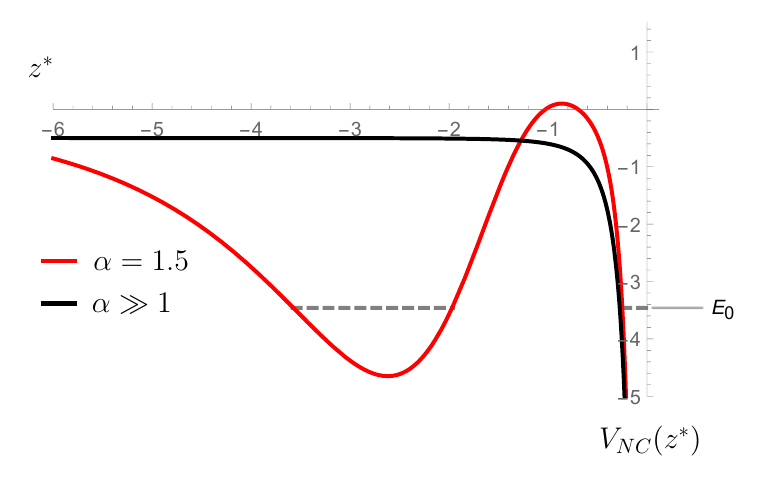}
         \caption{}
		\label{fig:newpot2b}
	\end{subfigure}
	\caption{\begin{small}(a) Far from extremality ($\nu =0.185$) with $\kappa = 0.61$, the NC effects develop two local minima, capturing the LLL. (b) The near extremal ($\nu =2.99$) black brane ($\kappa=\pi/2$) case. The normalizable mode $m^{2}L^{2}=-2$ in the C-case is unstable and no hair formation is present. However, turning the NC effects, this mode captures the LLL.\end{small}}   
    \label{fig:newpot2}
\end{figure}
Far away from the extremality, the NC effects can promote tunneling behavior for some values of the scalar field mass. To observe this effect, we take $\nu=1.675$ and $\kappa=0.084823$ with $m^{2}L^{2}=-1.44$ satisfying the BF bound (\ref{NCBF}). For the above values, we see in Fig.~\ref{fig:newPot2a} the formation of two potential wells one next to the other in the NC case ($\alpha\sim 1$) corresponding to the red curve. If the nearness parameter grows, we go back to the C-case, and one single local well captures the LLL. The same behavior takes place for all allowed massive modes, including the mode $m^2 L^2 =-2$, with a similar shape to the one depicted in Fig.~\ref{fig:newPot2a}. Transition amplitudes between the NC barrier would require the knowledge of approximate WKB solutions for the scalar field and the turning points for a given energy\footnote{See \cite{Cubrovic:2011xm} for instance, in the case of a Dirac field in a commutative geometry.}. Furthermore, it would be interesting to analyze these two minima in the context of a renormalization group flow between them \cite{Laia:2011wf}.

We stress that, in the opposite situation, i.e., when the configuration is far away from the extremality condition (i.e., $\nu \mapsto 0$), we are not ben able to find an increase in hair formation due to NC effects and the growing of bound states as the parameters are varied. The general behaviors rest in a simple potential well displacement in the $z^{*}$ coordinate relative to the C-case when the nearness parameter tends to one. To the interested reader, in the appendix \ref{pottoplay} the explicit expression of the Schrödinger potential (\ref{uniqueS}) in \emph{Mathematica} input code format is shown.

Finally, in Fig. \ref{fig:newpot2b} we go back to the extremal case and take $\kappa = \pi/2$, which corresponds to the purely magnetic black brane. For the normalizable value $m^{2}L^{2}=-2$, the potential does not capture the LLL state (depicted in the figure). Now, if we turn on the NC effects through the nearness parameter, a robust potential captures the LLL level, constituting a relevant novel result.

It is worth mentioning that in all cases described above, the NC background seems to be innocuous close to the UV boundary. In other words, the shape of all analyzed curves does not change in this region\footnote{The boundary region analysis requires approximating the star coordinate in the expression (\ref{zstarc}) close to $z=0$.}. This statement is in agreement with previous literature results concerning the holography theory on NC backgrounds (see \cite{Maeda:2009vf, Pramanik:2015eka}, for instance). We expect that the NC effects disappear in the asymptotically AdS boundary since this region is far away from the horizon radius and the NC scope. The next section addresses deeply the near-horizon and near-extremal structures, using one more suitable change of coordinates.

\section{Diving into the NC SAdS$_{2}$ black hole}\label{Spotential2}
The commutative dyonic solution also admits and AdS$_2\times\mathbb{R}^{2}$ Schwarzschild black brane structure (SAdS$_2$) \cite{Faulkner:2009wj}. This configuration allows us to explore the scalar hair in the near extremal (NE) and near-horizon (NH) regimes at finite small temperatures and, gives a coherent description of the density of superconducting states, confirming by the way, the results of the previous section. The NENH geometry is obtained with the knowledge of the extremal conditions that, in the NC version acquires contributions of $\alpha$ ($\theta$). Furthermore, when observing with more detail the behavior of the blackening factor comparison between the commutative and NC extremal cases, Fig.~\ref{fig:nhbkn}, we notice that in a neighborhood of the merged horizon, both functions are roughly similar. This fact invites us to develop the NC-extremal conditions to obtain a NENH effective IR scalar field equation at finite temperature in the NC geometry (called NCNENH from now on). Once we obtain the Schrödinger potential associated with the effective NCNENH scalar field equation, potential wells with finite energy are interpreted as bound states, dual to Cooper pairs, including the ground state (LLL). Again, we found new bound states relative to the commutative black hole explored in \cite{refId0} and whose existence is due solely to the NC effects, as we shall see.

Taking into account that the electromagnetic charges (\ref{reparam1}) have dimensions of length square, we can introduce a new parametrization in terms of a length scale $r_{*}$
\begin{equation}
    \mathcal{Q}=q r_{*}^2, \ \ \ \ \mathcal{B}=b r_{*}^2, \ \ \ \ \mathcal{F}=\left(q^2+b^2\right)r_{*}^4\equiv Fr_{*}^{4},
\end{equation}
with $q$ and $b$, dimensionless numbers. In this parametrization, and taking into account the nearness parameter (\ref{thebound0}), Hawking temperature can be cast in the form
\begin{equation}
    T\h=\dfrac{3r_{h}^{4}-Fr_{*}^{4}}{4\pi L^{2} r_{h}^{3}}-\dfrac{\sqrt{2}e^{-2\alpha^{2}}\left(r_{h}^{4}+Fr_{*}^{4}\right)\alpha^{3}}{r_{h}^{3}\pi L^{2}\tilde{\gamma}_{l}},
\end{equation}
and the NC extremality condition is the real positive root of $T_{H}=0$, namely
\begin{equation}
 r_{h}=r_{*}\left(\dfrac{F}{3}\right)^{1/4} \left(  1+\dfrac{32}{\dfrac{3\sqrt{2}\tilde{\gamma}_{l}}{\alpha^{3}}e^{2\alpha^{2}}-8}\right)^{1/4}\equiv \Tilde{r}.
\end{equation}
All the NC effects are captured on the last term of the above equation and, when $\alpha\mapsto\infty$, we recover the C-extremal condition \cite{refId0}. Furthermore, we can define two quantities that describe the proximity of the geometry (\ref{bknG}) to extremal condition and the horizon,
\begin{equation}\label{NearCoord0}
   \mbox{NE:}\quad r-\Tilde{r}=u\ll 1, \qquad \mbox{NH:} \quad r_{h}-\Tilde{r}=u_{0}\ll 1,
\end{equation}
followed by a series expansion up to second order around $(u,u_{0})=\mathbf{0}$. Finally, consider the following coordinate and the rescaling
\begin{equation}\label{NearCoord}
    u=\dfrac{\varepsilon l^{2}}{\zeta}\ll 1, \quad u_{0}=\dfrac{\varepsilon l^{2}}{\zeta_{0}}\ll 1,
\end{equation}
being $l$, the NC effective curvature radius that coincides exactly with (\ref{adsLeff}), as it should be. The above $\zeta$-coordinate capture the fact that the AdS$_2$ has a scaling limit $t\mapsto\varepsilon^{-1}\tau$ therefore, as long we take $\varepsilon\mapsto 0$ with $\zeta$ and $\tau$ finite \cite{Faulkner:2009wj}, it is possible to bring the full NC dyonic solution (\ref{blackeningsol}) into an NCNENH metric
\begin{equation}\label{AdS2bh}
    ds^{2}=\dfrac{l^{2}}{\zeta^{2}} \left(-f(\zeta)d\tau^{2}+\dfrac{d\zeta^{2}}{f(\zeta)}   \right)+\dfrac{\Tilde{r}^{2}}{L^{2}} d\vec{x}^{2},
\end{equation}
with blackening factor $f(\zeta)$ and Hawking temperature 
\begin{equation}\label{HawkingSAdS2}
    f(\zeta)=1-\left(1+\delta\right)\dfrac{\zeta^{2}}{\zeta_{0}^{2}}, \qquad \mathfrak{T}\h=\dfrac{\sqrt{1+\delta}}{2 \pi \zeta_{0}},
\end{equation}
where we have defined
\begin{equation}\label{deltap}
    \delta \equiv -\frac{8\sqrt{2} \alpha ^3 \left(4 \sqrt{2} \alpha^3+ e^{2 \alpha ^2} \left(4 \alpha ^2-3\right) \tilde{\gamma}_{l} \right)}{ e^{2 \alpha ^2} \tilde{\gamma}_{l}  \left(4\sqrt{2} \alpha^3  \left(4 \alpha ^2-1\right)+3  e^{2 \alpha ^2} \tilde{\gamma}_{l} \right)},
\end{equation}
that captures all the NC effects in such a way that $\delta\mapsto 0$ when $\alpha\mapsto\infty$. In the geometry (\ref{AdS2bh}), the horizon is located at $\zeta=\tfrac{\zeta_{0}}{\sqrt{1+\delta}}$ while the boundary is at $\zeta=0$. Besides, the scalar field equation (\ref{Req1}) takes the form\footnote{Also, due to covariant derivative, with the coordinate (\ref{NearCoord}) we need to consider the transformed expressions of the gauge potential components $A_{\tau}$ and $A_{x_{2}}$ and take the limit $\varepsilon\mapsto 0$. }
\begin{equation}\label{AdS2scalar}
    R''(\zeta)+\frac{f'(\zeta)}{f(\zeta)}R'(\zeta)+\dfrac{R(\zeta)}{\zeta^2 f(\zeta)}\left(\frac{q^2}{3\left(q^2+b^2\right)}\dfrac{\left(1-\zeta/\zeta_{0}\right)^2}{f(\zeta)}-l^2 m^2-\left(\frac{3}{q^2+b^2}\right)^{1/2}\frac{b\lambda_{n}}{3}\right)=0.
\end{equation}
According to this equation, in the AdS$_{2}$ boundary $\zeta=0$, the asymptotic solution  of $R(\zeta)$ has normalizable falloffs
\begin{equation}
    R(\zeta\mapsto 0)\approx c_{1}\zeta^{\frac{1}{2}\left(1+\eta\right)}+c_{2}\zeta^{\frac{1}{2}\left(1-\eta\right)},
\end{equation}
provided that
\begin{equation}
    \eta=\sqrt{1+4l^2m^2+\dfrac{4b\lambda_{n}}{\sqrt{3\left(q^2+b^2\right)}}},
\end{equation}
is real. From the above equation, an \emph{effective mass} arises in NCNENH for the perturbative scalar field
\begin{equation}
    M_{n}^2\equiv 4m^2+\dfrac{4b\lambda_{n}}{l^2\sqrt{3\left(q^2+b^2\right)}},
\end{equation}
acquiring discrete labels due to the Landau Levels associated with (\ref{gammaeq1}). 

In the BF window $(-9/4,-1/4)$ between AdS$_{4}$ and AdS$_{2}$ we can search for bound states of the scalar field equation. To achieve this aim, consider the Hawking temperature (\ref{HawkingSAdS2}), to express the blackening factor as
\begin{equation}\label{Hawkingfinal}
    f(\zeta)=1-\left(2 \pi \mathfrak{T}\h \zeta\right)^2.
\end{equation}
In the NCNENH geometry, the rescaling of (\ref{AdS2scalar}) given by  $R(\zeta )=s(\zeta )\, \mathcal{W} (\zeta )$\footnote{The same procedure that we implement to obtain (\ref{protoSch}).}, gives
\begin{equation}
    s(\zeta) =\dfrac{\sqrt{\zeta}}{\left(f\left(\zeta\right)\right)^{1/4}},
\end{equation}
and transforms (\ref{AdS2scalar}) into
\begin{equation}
-\zeta\sqrt{f\left(\zeta\right)}\partial_{\zeta}\left(\zeta\sqrt{f\left(\zeta\right)}\partial_{\zeta}\mathcal{W}\left(\zeta\right)\right)+V_{\Ncon}\left(\zeta\right)\mathcal{W}\left(\zeta\right)=E_{n}\mathcal{W}\left(\zeta\right).
\end{equation}
To write the unique Schrödinger potential, we define a star coordinate $\zeta_{*}$ such that $\partial_{\zeta_{*}}=\zeta\sqrt{f(\zeta)}\partial_{\zeta}$. Therefore
\begin{equation}\label{SchEq2}
-\partial_{\zeta_{*}}^{2}\mathcal{W}\left(\zeta_{*}\right)+ V\nh\left(\zeta_{*}\right)\mathcal{W}\left(\zeta_{*}\right)=E_{n}\mathcal{W}\left(\zeta_{*}\right),
\end{equation}
with energy spectrum
\begin{equation}
    E_{n}=-\dfrac{b}{\sqrt{3\left(q^{2}+b^{2}\right)}}\left(2 n+1\right).
\end{equation}
The physics is codified in the associated Schrödinger potential of (\ref{SchEq2})
\begin{equation}\label{Ads2potEq}
    \begin{aligned}
        V\nh(\zeta_{*}) &= -l^{2}m^{2}-\frac{1}{4}+\left(\pi\mathfrak{T}\h\zeta_{*}\right)^{2}\\
 &\quad + \frac{4 \pi ^2 \mathfrak{T}\h^2 }{3 \left(1-4 \pi ^2 \zeta_{*}^2 \mathfrak{T}\h^2\right)}\left(\frac{q^2 \left(\zeta_{*} -\frac{\sqrt{\delta +1}}{2 \pi  \mathfrak{T}\h}\right)^2}{(\delta +1) \left(b^2+q^2\right)}+3 \pi ^2 \zeta_{*}^{4}\mathfrak{T}\h^2\right)
    \end{aligned}
\end{equation}
and the NC effects are captured by the effective curvature $l$ (\ref{adsLeff}) and the $\delta$ parameter defined in (\ref{deltap}) which in turn depend both on $\alpha$, the nearness parameter. Moreover, the coordinate transformation
\begin{equation}
    \zeta_{*}=\int^{\zeta}\dfrac{d\zeta'}{\zeta'\sqrt{f(\zeta')}}=-\arctanh\sqrt{1-\left(2\pi\mathfrak{T}_{H}\zeta\right)^{2}},
\end{equation}
reveals that the boundary is now located at $\zeta_{*}\mapsto -\infty$ whereas the horizon is situated at $\zeta_{*}=0$.

\begin{figure}[h]
	\centering
	\begin{subfigure}[b]{.496\textwidth}
		\centering
		\includegraphics[width=.95\textwidth]{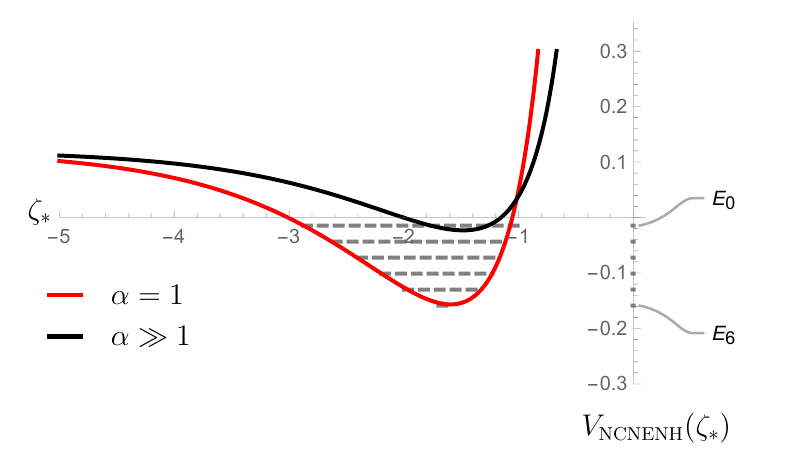}
		\label{fig:nceff1}
	\end{subfigure}
	\hfill     
	\begin{subfigure}[b]{.496\textwidth}
		\centering
		\includegraphics[width=.95\textwidth]{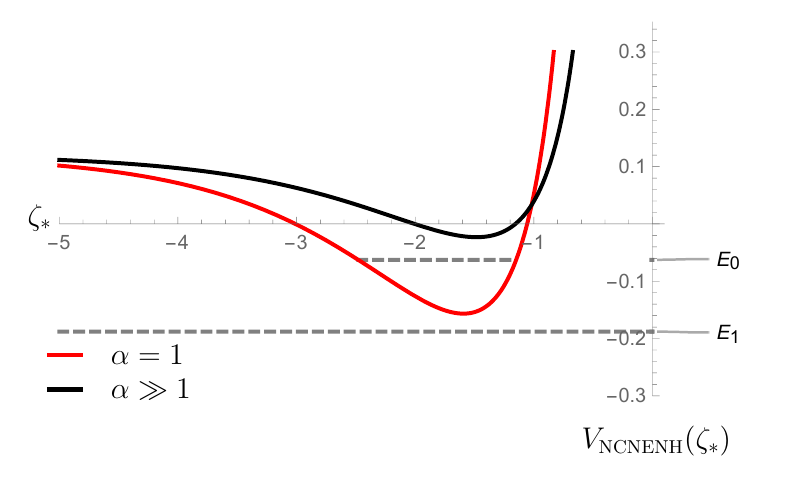}    
		\label{fig:nceff2}  
	\end{subfigure}
 \caption{\emph{Left panel}: Schrödinger potential in the C and NC cases where the former has the LLL $E_{0}$ bound state. NC effects increase the minimum of the potential, capturing more energy bands. \emph{Right panel}: In this case, the C-case has no bound states. NC effects allow for hair formation when $\alpha\mapsto 1$, capturing the LLL due to a a stepper potential well. We invite the reader to consult \cite{Faulkner:2009wj} for a similar construction in the context of the holographic Fermi surface.}
 \label{fig:hairAds2}
\end{figure}

In Fig.~\ref{fig:hairAds2}, we plot $V\nh$ vs. $\zeta_{*}$ for some values of $m$ in the BF window (\ref{NCBF}) and fixed electromagnetic charges. Both plots contain two graphs with different values of the nearness parameter, corresponding to $\alpha\gg 1$ which is the C-case (black curve), and $\alpha\sim 1$ (red curve) where NC effects are considered. Despite that we have a continuous nearness parameter, for clarity in the figures, we only depict two values of $\alpha$.

The left panel shows that the C-case has the ground state $E_{0}$ (corresponding to LLL) in the near horizon geometry ($\zeta_{*}\sim0$) and, descending $\alpha$ continuously, we observe the formation of a more pronounced potential well, favoring the hair condensate, capturing more bound states. The right panel plot, on the other hand, exhibits that modes of the scalar field in the commutative dyonic geometry do not promote hair formation for a given black hole parameters. Nonetheless, turning on the NC effects (descending the $\alpha$ values), the scalar hair continuously starts to form capturing the LLL\footnote{To the interested reader, in the appendix \ref{pottoplay} the explicit expression of the Schrödinger potential (\ref{Ads2potEq}) in \emph{Mathematica} input code format is shown.}.

The above-described plots are in agreement with the results obtained when considering the full scalar field equation, showing us that bound states can exist in the IR region. Such a result allows us to confirm the existence of a superconducting state solely by the NC effects in the IR regime, compatible with expectations from seminal works \cite{Gubser:2008px, Faulkner:2009wj}. These results constitute the principal contribution of this report and, as far as the authors know, have not been addressed in previous literature.

\section{Conclusions and further investigations}
In this work, we study quite exhaustively a $3+1$ dimensional NC-charged black hole with a negative cosmological constant. The background represents a confined box whose NC effects are captured only by the Gaussian distribution of the black hole mass and the electromagnetic charges remain commutative. Despite this setup does not cure the essential singularity and constitutes a particular case of more general NC constructions (Maxwell fields also NC), it allows us to explore the effects of this background over a perturbative non-backreacting charged scalar field. In the context of the holographic description of superconductivity, our framework is dual to a minimal type-II superconductor where vacuum expectation values for the order parameter are described by the boundary values of the charged scalar field. The adjective \emph{minimal} refers to the fact that the Maxwell field is fixed and does not interact dynamically with the scalar field (unlike the celebrated holographic Abelian-Higgs model). However, this model certainly does a well qualitative description of the upper critical magnetic fields; regions below which a vortex lattice structure starts to form, surrounded by a supercurrent density. Even more, our NC dyonic black hole possesses an AdS$_2\times\mathbb{R}^{2}$ topology near the horizon and then, is capable of describing the infrared regime of the superconductor.

The novelty of this construction rests in the semi-analytical description of the NC effects and the continuous limit to the C-case. Some of the main quantities regarding the holographic description of superconductivity in previous literature are enriched by the NC effects.

We also study the relevant thermodynamical quantities, keeping track of the NC effects by the use of a defined nearness parameter that accounts for the size of the horizon radius relative to the NC original parameter $\theta$. We found that the NC Nicolini mass distribution satisfies the first law of Thermodynamics using Brown-York techniques. The stability of the background is studied with an equation of state, revealing that, as long as the nearness parameter remains above the cutoff, the bulk is stable. All thermodynamic variables reduce to the commutative dyonic black hole ones when the NC nearness parameter continuously grows. 

The aforementioned nearness parameter has a cutoff that naturally emerges in the functional dependency of the Ricci scalar curvature and the effective curvature radius in the near horizon geometry. We interpret this bound as the minimal size for the horizon radius that the NC background supports. Therefore, close to this bound, the NC effects are more intense. At this point, our background acquires the most important feature, namely: the IR regime close to the horizon. The scalar field must \emph{feel} the NC spacetime with more amplitude than in the boundary region. We found, by using the scalar field Schrödinger potential that, indeed, close to the horizon, the NC spacetime acts in favor of hair formation. Thanks to this continuous nearness parameter, we can control the NC effects by a fine-tuning procedure and we found also that, for certain values of the parameters of the background and the scalar field mass, the commutative dyonic bulk can not retain hair but, turning on the NC effects, a robust hair continuously starts to form, constituting the principal result of this report.

Using the so-called matching method, we also obtain a semi-analytical phase space of the superconductor, the thermal behavior of the upper critical magnetic field controlled continuously by the nearness parameter. In the canonical ensemble case, we found a widening in the shape of the magnetic field thermal behavior whereas, for the grand canonical one, a major amplitude close to zero temperature. Both analyzed cases, are relative to the commutative case previously reported.

To reinforce the existence of hair solely by the NC effects, we explored the NC near-extremal and near-horizon (NCNENH) geometry, adopting a Schwarzschild AdS$_2\times\mathbb{R}^{2}$ structure and Hawking temperature with NC contributions. The Schrödinger potential associated with the effective scalar field in this regime confirms the existence of bound states (density of Cooper pairs) that the commutative NENH geometry forbids.

In summary, the new results of this work rest in the following three general statements:
\begin{itemize}
    \item We revealed the near-horizon AdS$_{2}\times\mathbb{R}^{2}$ topology of the NC black hole background that describes the IR regime of type-II superconductors, a regime that has been previously underestimated in the literature, which usually focuses on boundary UV studies.
     \item We decoupled the scalar field from the analytic Einstein-Maxwell AdS background, in contrast to the Higgs-Abelian model or the full back-reacted Einstein-Maxwell-scalar configuration that allows for just numerical approaches.
    \item We considered a perturbative scalar field on a near-horizon and near-extremal charged black hole to study the enhancement of bound states completely driven by NC effects in the steeper AdS$_{2}$ throat region comparable to the commutative case.
\end{itemize}
In particular, a more detailed account of the new results reads:
\begin{itemize}
    \item We defined a novel nearness parameter $\alpha$ with a natural cutoff that controls the intensity of NC effects, particularly in the near-horizon region, and continuously approaches the commutative limit.
    \item We studied the thermodynamical properties of the AdS gravitational background compatible with the fundamental first law in terms of the $\alpha$ and showed that it corresponds to a stable configuration for values above its cutoff. 
    \item We expressed the near-horizon effective curvature as a function of $\alpha$ that parametrizes the inward gravitational pulling of the AdS$_{2}$ throat.
    \item We obtained a new effective scalar field mass that respects the Breitenlohner-Friedmann bound in terms of $\alpha$.
    \item We expressed the upper critical magnetic field as a function of the critical temperature with continuous dependency on $\alpha$ for regions where the Abrikosov vortex lattice emerges. 
    \item We obtained the effective Schrödinger potential for the dynamics of the scalar field and revealed three novel effects continuously driven by $\alpha$: the enhancement in the number of bound states responsible for superconductivity, a tunneling effect of the Lowest Landau Level, and the existence of purely NC bound states that eventually could experimentally uncover the NC nature of certain dual superconductors.
    \item We used an analytic approach borrowed from emergent quantum criticality studies to get access to the near-horizon and near-extremal background and confirm the existence of the above-mentioned NC bound states. 
    \end{itemize}

It is important to mention that the novel results found in this work and listed above could potentially be improved in more general NC solutions, such as those with NC electromagnetic charges or in Abelian-Higgs models. We leave it to future work to extend these results to more general gravitational backgrounds. An interesting aspect that also we could explore in the future was found when we studied the Schrödinger potential. We show in this report that NC effects can generate two local minima. It would be very interesting to explore the holographic renormalization flow between these two fixed points, in a similar fashion studied in \cite{Faulkner:2010jy, Laia:2011wf}.

The program of holographic superconductivity has recently been reborn due to the imperative need to describe the Meissner effect. In standard holographic superconductivity, there is no Meissner effect because the Maxwell source is not dynamical. In this regard, it was shown in \cite{Xia:2021jzh} that, by turning on the dynamics of the boundary expansion coefficients of the Maxwell and the scalar field, a numerical curve for the lower critical magnetic field was obtained, extending the phase space to the regime below which the Meissner effect is present (see also \cite{Domenech:2010nf}). Besides, an important effort in the same direction was developed in \cite{Natsuume:2024ril} where the author obtained a holographic dual of the Ginzburg-Landau mean-field theory, computing its free parameters with gravitational duals.

On the other hand, recall the theoretical connection between the smearing out mass distributions with the NC fluctuations of the spacetime manifold \cite{Nicolini:2008aj, Banerjee:2009xx} and that the NC effects can also be implemented perturbatively via the gauge-invariant Seiberg-Witten map \cite{Seiberg:1999vs}. One of the Euler-Lagrange equations of the Ginzburg-Landau free energy is precisely the London equation that accounts for the Meissner effects and, the NC description on this equation using the Seiberg-Witten map has been studied in \cite{Martinez-Carbajal:2021kfp}, resulting in an effective non-local London penetration length. Guided by the above works, the current authors are interested in merging, in a single and self-consistent framework, the NC Seiberg-Witten version of the Ginzburg-Landau mean-field theory with an inspired NC holographic dyonic black hole superconductivity, studied in this report. 
\section*{Acknowledgments}
\begin{small} 
All authors are grateful to Jhony A. Herrera-Mendoza, Marco Maceda, Mehrab Momenia, and Uriel Noriega Cornelio for enriching discussions. MCL especially thanks Yasmin Vázquez, Fernando Ochoa and the financial assistance provided by a CONAHCYT postdoctoral grant with CVU 393605 as well as SNII. AHA acknowledges support from SNII and a VIEP grant. DMC acknowledges financial support from TecNM and SNII. The work of SPL was partially supported by CONAHCYT grant A1-S-22886 and DGAPA-UNAM grant IN116823. SPL is thankful to Giorgos Anastasiou and Adolfo Ibáñez University for their hospitality. 
\end{small}

\textbf{Data Availability Statement}: No Data associated in the manuscript.
\appendix
\section{NC quantities}\label{coefficients}
The functions $(a_{i},b_{i})$ of the phase space expressions (\ref{phaseS}) (in the canonical ensemble) for a general scalar field mass, despite we take at the end, the value $L^{2}m^{2}=-2$. For the $a_{i}$ coefficients we have

\begin{small}
\begin{align}
   a_{1}&=-\dfrac{4 e^{4 \alpha ^2} (z_{m}-1)^2 \tilde{\gamma}_{l}^2}{\left(8 \sqrt{2}\alpha ^3-6e^{2 \alpha ^2}  \tilde{\gamma}_{l}\right)^2}=-a_2,\\
   a_{3} &=-\dfrac{ 4e^{2 \alpha ^2}(z_{m}-1)\tilde{\gamma}_{l}}{\left(8\sqrt{2} \alpha ^3 -6 e^{2 \alpha ^2}\tilde{\gamma}_{l}\right)^2} \Biggl[ \left(8\sqrt{2} \alpha ^5 -e^{2 \alpha ^2}L^2 m^2\tilde{\gamma}_{l}\right)\left( z_m-1\right)-8\sqrt{2} \alpha ^3 +6 e^{2 \alpha ^2}\tilde{\gamma}_{l}^2\Biggr],\\
   a_{4} &= \dfrac{1}{\left(8\sqrt{2} \alpha ^3 -6 e^{2 \alpha ^2}\tilde{\gamma}_{l}\right)^2} \Biggl[ \left(8\sqrt{2} \alpha ^3 -6 e^{2 \alpha ^2}\tilde{\gamma}_{l}\right)^2\nonumber\\
   &\quad
   +e^{2 \alpha ^2}\tilde{\gamma}_{l} L^2 m^2 \left(z_m-1\right) \left(\left(8\sqrt{2} \alpha ^3 -6 e^{2 \alpha ^2}\tilde{\gamma}_{l}\right)\left(z_m-3\right)+\left(16\sqrt{2} \alpha ^5 -e^{2 \alpha ^2}L^2m^2\tilde{\gamma}_{l}\right)\left(z_m-1\right)\right)
   \Biggr].
  \end{align}
\end{small}
Regarding the $b_{i}$ coefficients,

\begin{small}
\begin{align}
   b_{1} &=-\dfrac{8 e^{4 \alpha ^2} (z_{m}-1) \tilde{\gamma}_{l}^2}{\left(8 \sqrt{2}\alpha ^3-6e^{2 \alpha ^2} \tilde{\gamma}_{l}\right)^2}=-b_{2},\\
   b_{3} &=\dfrac{2 e^{2 \alpha ^2}\tilde{\gamma}_{l}}{\left(8\sqrt{2} \alpha ^3 -6 e^{2 \alpha ^2}\tilde{\gamma}_{l}\right)^2} \Biggl[  4\sqrt{2}\alpha^3-3e^{2 \alpha ^2}\tilde{\gamma}_{l}-\left( 8\sqrt{2}\alpha^5- e^{2 \alpha ^2}L^2 m^2\tilde{\gamma}_{l} \right)(z_m-1)\Biggr],\\
   b_{4} &=-\dfrac{2 L^2 m^2 e^{2 \alpha ^2}\tilde{\gamma}_{l}}{\left(8\sqrt{2} \alpha ^3 -6 e^{2 \alpha ^2}\tilde{\gamma}_{l}\right)^2}
   \Biggl[\left(8\sqrt{2} \alpha ^3 -6 e^{2 \alpha ^2}\tilde{\gamma}_{l}\right)(z_m-2)+ \left(16\sqrt{2} \alpha ^5 - e^{2 \alpha ^2}L^2m^2\tilde{\gamma}_{l}\right)(z_m-1)\Biggr].
\end{align}
\end{small}

\section{NC Schrödinger potential}\label{pottoplay}
The AdS$_{2}$ NCNENH Schrödinger potential (\ref{Ads2potEq})

\begin{tiny}
\begin{verbatim}
Manipulate[
 Plot[{1/12 (-3 - (12 l^2 m)/(
      1 + (16 Sqrt[2] \[Alpha]^3 (-1 + \[Alpha]^2))/(
       3 (4 Sqrt[2] \[Alpha]^3 + 
          E^(2 \[Alpha]^2) Gamma[3/2, 0, 2 \[Alpha]^2]))) + 
      3 Sech[z]^2 + 
      Coth[z]^2 (3 Sech[z]^4 + (
         4 v Cos[k]^2 (Sqrt[
            1 + (32 \[Alpha]^3 (6 Sqrt[2] \[Alpha] + 
                E^(2 \[Alpha]^2)
                  Sqrt[\[Pi]] (-3 + 4 \[Alpha]^2) Erf[
                  Sqrt[2] \[Alpha]]))/((2 Sqrt[2] \[Alpha] - 
                E^(2 \[Alpha]^2) Sqrt[\[Pi]]
                  Erf[Sqrt[2] \[Alpha]]) (4 \[Alpha] (-3 - 
                   4 \[Alpha]^2 + 16 \[Alpha]^4) + 
                3 E^(2 \[Alpha]^2) Sqrt[2 \[Pi]]
                  Erf[Sqrt[2] \[Alpha]]))] - Sqrt[Sech[z]^2])^2)/(
         v + (32 v \[Alpha]^3 (6 Sqrt[2] \[Alpha] + 
             E^(2 \[Alpha]^2)
               Sqrt[\[Pi]] (-3 + 4 \[Alpha]^2) Erf[
               Sqrt[2] \[Alpha]]))/((2 Sqrt[2] \[Alpha] - 
             E^(2 \[Alpha]^2) Sqrt[\[Pi]]
               Erf[Sqrt[2] \[Alpha]]) (4 \[Alpha] (-3 - 
                4 \[Alpha]^2 + 16 \[Alpha]^4) + 
             3 E^(2 \[Alpha]^2) Sqrt[2 \[Pi]]
               Erf[Sqrt[2] \[Alpha]]))))), 
   Table[-(((1 + 2 n) Sin[k])/Sqrt[3]), {n, 0, 20, 1}]}
  , {z, -5, 0}, PlotRange -> {-0.5, 0.5}], {l, 0.001, 
  1}, {k, -(\[Pi]/2), \[Pi]/2}, {m, -9/4, -1/4}, {\[Alpha], 1, 7}]
\end{verbatim}
\end{tiny}

Regarding the curves shown in chapter \ref{Spotential1}
\begin{tiny}
\begin{verbatim}
   Manipulate[
 Plot[{Vpot[z, m, v, k, \[Alpha]], 
   Table[-2 (2*n + 1)* Sqrt[v] Sin[k], {n, 0, 10, 1}]}, {z, -4, 0}, 
  PlotRange -> {-6, 1}, 
  WorkingPrecision -> 100], {L, {1}}, {a, {1}}, {k, 0.001, \[Pi]/
  2}, {v, 0.001, 2.99}, {m, -(9/4), -(1/4)}, {\[Alpha], 1, 10}] 
\end{verbatim}
\end{tiny}
Being the potential (\ref{uniqueS})
\newpage
\begin{multicols}{2}
\begin{tiny}
\begin{verbatim}
Vpot[z_, m_, v_, k_, \[Alpha]_] := (
  16384 Sqrt[2]
    E^(-((128 \[Alpha]^2 Gamma[3/2, 0, 
      2 \[Alpha]^2]^2)/(2 Sqrt[2] E^(-2 \[Alpha]^2)
        z^2 \[Alpha] (3 - v + 4 (1 + v) \[Alpha]^2) + 
      Sqrt[\[Pi]] (-3 + v) z^2 Erf[Sqrt[2] \[Alpha]] + 
      8 Gamma[3/2, 0, 2 \[Alpha]^2])^2)) \[Alpha]^5 Gamma[3/2, 0, 
    2 \[Alpha]^2]^3)/(2 Sqrt[2] E^(-2 \[Alpha]^2)
      z^2 \[Alpha] (3 - v + 4 (1 + v) \[Alpha]^2) + 
    Sqrt[\[Pi]] (-3 + v) z^2 Erf[Sqrt[2] \[Alpha]] + 
    8 Gamma[3/2, 0, 2 \[Alpha]^2])^4 + (
  16384 Sqrt[2]
    E^(-((128 \[Alpha]^2 Gamma[3/2, 0, 
      2 \[Alpha]^2]^2)/(2 Sqrt[2] E^(-2 \[Alpha]^2)
        z^2 \[Alpha] (3 - v + 4 (1 + v) \[Alpha]^2) + 
      Sqrt[\[Pi]] (-3 + v) z^2 Erf[Sqrt[2] \[Alpha]] + 
      8 Gamma[3/2, 0, 2 \[Alpha]^2])^2))
    v \[Alpha]^5 Gamma[3/2, 0, 
    2 \[Alpha]^2]^3)/(2 Sqrt[2] E^(-2 \[Alpha]^2)
      z^2 \[Alpha] (3 - v + 4 (1 + v) \[Alpha]^2) + 
    Sqrt[\[Pi]] (-3 + v) z^2 Erf[Sqrt[2] \[Alpha]] + 
    8 Gamma[3/2, 0, 2 \[Alpha]^2])^4 - (
  128 Sqrt[2]
    E^(-((
    128 \[Alpha]^2 Gamma[3/2, 0, 
      2 \[Alpha]^2]^2)/(2 Sqrt[2] E^(-2 \[Alpha]^2)
        z^2 \[Alpha] (3 - v + 4 (1 + v) \[Alpha]^2) + 
      Sqrt[\[Pi]] (-3 + v) z^2 Erf[Sqrt[2] \[Alpha]] + 
      8 Gamma[3/2, 0, 2 \[Alpha]^2])^2)) \[Alpha]^3 Gamma[3/2, 0, 
    2 \[Alpha]^2])/(2 Sqrt[2] E^(-2 \[Alpha]^2)
      z^2 \[Alpha] (3 - v + 4 (1 + v) \[Alpha]^2) + 
    Sqrt[\[Pi]] (-3 + v) z^2 Erf[Sqrt[2] \[Alpha]] + 
    8 Gamma[3/2, 0, 2 \[Alpha]^2])^2 - (
  128 Sqrt[2]
    E^(-((128 \[Alpha]^2 Gamma[3/2, 0, 
      2 \[Alpha]^2]^2)/(2 Sqrt[2] E^(-2 \[Alpha]^2)
        z^2 \[Alpha] (3 - v + 4 (1 + v) \[Alpha]^2) + 
      Sqrt[\[Pi]] (-3 + v) z^2 Erf[Sqrt[2] \[Alpha]] + 
      8 Gamma[3/2, 0, 2 \[Alpha]^2])^2))
    v \[Alpha]^3 Gamma[3/2, 0, 
    2 \[Alpha]^2])/(2 Sqrt[2] E^(-2 \[Alpha]^2)
      z^2 \[Alpha] (3 - v + 4 (1 + v) \[Alpha]^2) + 
    Sqrt[\[Pi]] (-3 + v) z^2 Erf[Sqrt[2] \[Alpha]] + 
    8 Gamma[3/2, 0, 2 \[Alpha]^2])^2 + (
  128 Gamma[3/2, 0, 
    2 \[Alpha]^2]^2)/(2 Sqrt[2] E^(-2 \[Alpha]^2)
      z^2 \[Alpha] (3 - v + 4 (1 + v) \[Alpha]^2) + 
    Sqrt[\[Pi]] (-3 + v) z^2 Erf[Sqrt[2] \[Alpha]] + 
    8 Gamma[3/2, 0, 2 \[Alpha]^2])^2 + (
  64 m Gamma[3/2, 0, 
    2 \[Alpha]^2]^2)/(2 Sqrt[2] E^(-2 \[Alpha]^2)
      z^2 \[Alpha] (3 - v + 4 (1 + v) \[Alpha]^2) + 
    Sqrt[\[Pi]] (-3 + v) z^2 Erf[Sqrt[2] \[Alpha]] + 
    8 Gamma[3/2, 0, 2 \[Alpha]^2])^2 + (
  v (2 Sqrt[2] E^(-2 \[Alpha]^2)
       z^2 \[Alpha] (3 - v + 4 (1 + v) \[Alpha]^2) + 
     Sqrt[\[Pi]] (-3 + v) z^2 Erf[Sqrt[2] \[Alpha]] + 
     8 Gamma[3/2, 0, 2 \[Alpha]^2])^2)/(
  64 Gamma[3/2, 0, 
    2 \[Alpha]^2]^2) - ((2 Sqrt[2] E^(-2 \[Alpha]^2)
       z^2 \[Alpha] (3 - v + 4 (1 + v) \[Alpha]^2) + 
     Sqrt[\[Pi]] (-3 + v) z^2 Erf[Sqrt[2] \[Alpha]] + 
     8 Gamma[3/2, 0, 2 \[Alpha]^2]) Gamma[3/2, 0, (
    128 \[Alpha]^2 Gamma[3/2, 0, 
      2 \[Alpha]^2]^2)/(2 Sqrt[2] E^(-2 \[Alpha]^2)
        z^2 \[Alpha] (3 - v + 4 (1 + v) \[Alpha]^2) + 
      Sqrt[\[Pi]] (-3 + v) z^2 Erf[Sqrt[2] \[Alpha]] + 
      8 Gamma[3/2, 0, 2 \[Alpha]^2])^2])/(
  16 Gamma[3/2, 0, 2 \[Alpha]^2]^2) - (
  v (2 Sqrt[2] E^(-2 \[Alpha]^2)
       z^2 \[Alpha] (3 - v + 4 (1 + v) \[Alpha]^2) + 
     Sqrt[\[Pi]] (-3 + v) z^2 Erf[Sqrt[2] \[Alpha]] + 
     8 Gamma[3/2, 0, 2 \[Alpha]^2]) Gamma[3/2, 0, (
    128 \[Alpha]^2 Gamma[3/2, 0, 
      2 \[Alpha]^2]^2)/(2 Sqrt[2] E^(-2 \[Alpha]^2)
        z^2 \[Alpha] (3 - v + 4 (1 + v) \[Alpha]^2) + 
      Sqrt[\[Pi]] (-3 + v) z^2 Erf[Sqrt[2] \[Alpha]] + 
      8 Gamma[3/2, 0, 2 \[Alpha]^2])^2])/(
  16 Gamma[3/2, 0, 
    2 \[Alpha]^2]^2) - ((
     E^(-4 \[Alpha]^2)
       v z^4 Cos[
       k]^2 (2 Sqrt[
         2] \[Alpha] (-3 + v - 4 \[Alpha]^2 - 4 v \[Alpha]^2) + 
        E^(2 \[Alpha]^2)
          Sqrt[\[Pi]] (3 - v) Erf[Sqrt[2] \[Alpha]])^2)/
     Gamma[3/2, 0, 
      2 \[Alpha]^2]^2 + (E^(-((
         256 \[Alpha]^2 Gamma[3/2, 0, 
           2 \[Alpha]^2]^2)/(2 Sqrt[2] E^(-2 \[Alpha]^2)
             z^2 \[Alpha] (3 - v + 4 (1 + v) \[Alpha]^2) + 
           Sqrt[\[Pi]] (-3 + v) z^2 Erf[Sqrt[2] \[Alpha]] + 
           8 Gamma[3/2, 0, 2 \[Alpha]^2])^2)) (E^((
           128 \[Alpha]^2 Gamma[3/2, 0, 
             2 \[Alpha]^2]^2)/(2 Sqrt[2] E^(-2 \[Alpha]^2)
               z^2 \[Alpha] (3 - v + 4 (1 + v) \[Alpha]^2) + 
             Sqrt[\[Pi]] (-3 + v) z^2 Erf[Sqrt[2] \[Alpha]] + 
             8 Gamma[3/2, 0, 2 \[Alpha]^2])^2)
            v (2 Sqrt[2] E^(-2 \[Alpha]^2)
               z^2 \[Alpha] (3 - v + 4 (1 + v) \[Alpha]^2) + 
             Sqrt[\[Pi]] (-3 + v) z^2 Erf[Sqrt[2] \[Alpha]] + 
             8 Gamma[3/2, 0, 2 \[Alpha]^2])^4 + 
          2 (1 + v) (2048 Sqrt[
              2] \[Alpha]^3 Gamma[3/2, 0, 2 \[Alpha]^2]^3 - 
             3 E^((128 \[Alpha]^2 Gamma[3/2, 0, 
                2 \[Alpha]^2]^2)/(2 Sqrt[2] E^(-2 \[Alpha]^2)
                  z^2 \[Alpha] (3 - v + 4 (1 + v) \[Alpha]^2) + 
                Sqrt[\[Pi]] (-3 + v) z^2 Erf[Sqrt[2] \[Alpha]] + 
                8 Gamma[3/2, 0, 2 \[Alpha]^2])^2) (2 Sqrt[2]
                  E^(-2 \[Alpha]^2)
                  z^2 \[Alpha] (3 - v + 4 (1 + v) \[Alpha]^2) + 
                Sqrt[\[Pi]] (-3 + v) z^2 Erf[Sqrt[2] \[Alpha]] + 
                8 Gamma[3/2, 0, 2 \[Alpha]^2])^3 Gamma[3/2, 0, (
               128 \[Alpha]^2 Gamma[3/2, 0, 
                 2 \[Alpha]^2]^2)/(2 Sqrt[2] E^(-2 \[Alpha]^2)
                   z^2 \[Alpha] (3 - v + 4 (1 + v) \[Alpha]^2) + 
                 Sqrt[\[Pi]] (-3 + v) z^2 Erf[Sqrt[2] \[Alpha]] + 
                 8 Gamma[3/2, 0, 2 \[Alpha]^2])^2]))^2)/(16384 Gamma[
         3/2, 0, 2 \[Alpha]^2]^6 (2 Sqrt[2] E^(-2 \[Alpha]^2)
            z^2 \[Alpha] (3 - v + 4 (1 + v) \[Alpha]^2) + 
          Sqrt[\[Pi]] (-3 + v) z^2 Erf[Sqrt[2] \[Alpha]] + 
          8 Gamma[3/2, 0, 2 \[Alpha]^2])^2))/(16 (1 + (
       v (2 Sqrt[2] E^(-2 \[Alpha]^2)
            z^2 \[Alpha] (3 - v + 4 (1 + v) \[Alpha]^2) + 
          Sqrt[\[Pi]] (-3 + v) z^2 Erf[Sqrt[2] \[Alpha]] + 
          8 Gamma[3/2, 0, 2 \[Alpha]^2])^4)/(
       4096 Gamma[3/2, 0, 
         2 \[Alpha]^2]^4) - ((1 + 
          v) (2 Sqrt[2] E^(-2 \[Alpha]^2)
            z^2 \[Alpha] (3 - v + 4 (1 + v) \[Alpha]^2) + 
          Sqrt[\[Pi]] (-3 + v) z^2 Erf[Sqrt[2] \[Alpha]] + 
          8 Gamma[3/2, 0, 2 \[Alpha]^2])^3 Gamma[3/2, 0, (
         128 \[Alpha]^2 Gamma[3/2, 0, 
           2 \[Alpha]^2]^2)/(2 Sqrt[2] E^(-2 \[Alpha]^2)
             z^2 \[Alpha] (3 - v + 4 (1 + v) \[Alpha]^2) + 
           Sqrt[\[Pi]] (-3 + v) z^2 Erf[Sqrt[2] \[Alpha]] + 
           8 Gamma[3/2, 0, 2 \[Alpha]^2])^2])/(
       512 Gamma[3/2, 0, 2 \[Alpha]^2]^4)))
\end{verbatim}
\end{tiny}
\end{multicols}
\bibliography{references}

\providecommand{\href}[2]{#2}\begingroup\raggedright\begin{thebibliography}{10}

\bibitem{Hartnoll:2008vx}
S.~A. Hartnoll, C.~P. Herzog, and G.~T. Horowitz, ``{Building a Holographic Superconductor},'' {\em Phys. Rev. Lett.} {\bf 101} (2008) 031601, \href{http://www.arXiv.org/abs/0803.3295}{{\tt 0803.3295}}.

\bibitem{Hartnoll:2008kx}
S.~A. Hartnoll, C.~P. Herzog, and G.~T. Horowitz, ``{Holographic Superconductors},'' {\em JHEP} {\bf 12} (2008) 015, \href{http://www.arXiv.org/abs/0810.1563}{{\tt 0810.1563}}.

\bibitem{Maldacena:1997re}
J.~M. Maldacena, ``{The Large N limit of superconformal field theories and supergravity},'' {\em Adv. Theor. Math. Phys.} {\bf 2} (1998) 231--252, \href{http://www.arXiv.org/abs/hep-th/9711200}{{\tt hep-th/9711200}}.

\bibitem{zaanen_liu_sun_schalm_2015}
J.~Zaanen, Y.~Liu, Y.-W. Sun, and K.~Schalm, {\em Holographic Duality in Condensed Matter Physics}.
\newblock Cambridge University Press, 2015.

\bibitem{PhysRevD.106.L081902}
J.~A. Herrera-Mendoza, D.~F. Higuita-Borja, J.~A. M\'{e}ndez-Zavaleta, A.~Herrera-Aguilar, and F.~P\'{e}rez-Rodr\'{i}guez, ``{Vortex structure deformation of rotating Lifshitz holographic superconductors},'' {\em Phys. Rev. D} {\bf 106} (2022), no.~8, L081902, \href{http://www.arXiv.org/abs/2208.05988}{{\tt 2208.05988}}.

\bibitem{Herrera-Mendoza:2024vfj}
J.~A. Herrera-Mendoza, A.~Herrera-Aguilar, D.~F. Higuita-Borja, J.~A. M\'endez-Zavaleta, F.~P\'erez-Rodr\'\i{}guez, and J.-X. Yin, ``{Effects of rotation and anisotropy on the properties of type-II holographic superconductors},'' \href{http://www.arXiv.org/abs/2406.05351}{{\tt 2406.05351}}.

\bibitem{Nakano:2008xc}
E.~Nakano and W.-Y. Wen, ``{Critical magnetic field in a holographic superconductor},'' {\em Phys. Rev. D} {\bf 78} (2008) 046004, \href{http://www.arXiv.org/abs/0804.3180}{{\tt 0804.3180}}.

\bibitem{Gregory:2009fj}
R.~Gregory, S.~Kanno, and J.~Soda, ``{Holographic Superconductors with Higher Curvature Corrections},'' {\em JHEP} {\bf 10} (2009) 010, \href{http://www.arXiv.org/abs/0907.3203}{{\tt 0907.3203}}.

\bibitem{Gubser:2008px}
S.~S. Gubser, ``{Breaking an Abelian gauge symmetry near a black hole horizon},'' {\em Phys. Rev. D} {\bf 78} (2008) 065034, \href{http://www.arXiv.org/abs/0801.2977}{{\tt 0801.2977}}.

\bibitem{Hertog:2006rr}
T.~Hertog, ``{Towards a Novel no-hair Theorem for Black Holes},'' {\em Phys. Rev. D} {\bf 74} (2006) 084008, \href{http://www.arXiv.org/abs/gr-qc/0608075}{{\tt gr-qc/0608075}}.

\bibitem{Montull:2009fe}
M.~Montull, A.~Pomarol, and P.~J. Silva, ``{The Holographic Superconductor Vortex},'' {\em Phys. Rev. Lett.} {\bf 103} (2009) 091601, \href{http://www.arXiv.org/abs/0906.2396}{{\tt 0906.2396}}.

\bibitem{Maeda:2009vf}
K.~Maeda, M.~Natsuume, and T.~Okamura, ``{Vortex lattice for a holographic superconductor},'' {\em Phys. Rev. D} {\bf 81} (2010) 026002, \href{http://www.arXiv.org/abs/0910.4475}{{\tt 0910.4475}}.

\bibitem{Albash:2009iq}
T.~Albash and C.~V. Johnson, ``{Vortex and Droplet Engineering in Holographic Superconductors},'' {\em Phys. Rev. D} {\bf 80} (2009) 126009, \href{http://www.arXiv.org/abs/0906.1795}{{\tt 0906.1795}}.

\bibitem{Adams:2012pj}
A.~Adams, P.~M. Chesler, and H.~Liu, ``{Holographic Vortex Liquids and Superfluid Turbulence},'' {\em Science} {\bf 341} (2013) 368--372, \href{http://www.arXiv.org/abs/1212.0281}{{\tt 1212.0281}}.

\bibitem{Srivastav:2023qof}
A.~Srivastav and S.~Gangopadhyay, ``{Vortices in a rotating holographic superfluid with Lifshitz scaling},'' {\em Phys. Rev. D} {\bf 107} (2023), no.~8, 086005, \href{http://www.arXiv.org/abs/2302.01030}{{\tt 2302.01030}}.

\bibitem{Xia:2019eje}
C.-Y. Xia, H.-B. Zeng, H.-Q. Zhang, Z.-Y. Nie, Y.~Tian, and X.~Li, ``{Vortex Lattice in a Rotating Holographic Superfluid},'' {\em Phys. Rev. D} {\bf 100} (2019), no.~6, 061901, \href{http://www.arXiv.org/abs/1904.10925}{{\tt 1904.10925}}.

\bibitem{Donos:2020viz}
A.~Donos, J.~P. Gauntlett, and C.~Pantelidou, ``{Holographic Abrikosov Lattices},'' {\em JHEP} {\bf 07} (2020) 095, \href{http://www.arXiv.org/abs/2001.11510}{{\tt 2001.11510}}.

\bibitem{Romans:1991nq}
L.~J. Romans, ``{Supersymmetric, cold and lukewarm black holes in cosmological Einstein-Maxwell theory},'' {\em Nucl. Phys. B} {\bf 383} (1992) 395--415, \href{http://www.arXiv.org/abs/hep-th/9203018}{{\tt hep-th/9203018}}.

\bibitem{Hartnoll:2007ai}
S.~A. Hartnoll and P.~Kovtun, ``{Hall conductivity from dyonic black holes},'' {\em Phys. Rev. D} {\bf 76} (2007) 066001, \href{http://www.arXiv.org/abs/0704.1160}{{\tt 0704.1160}}.

\bibitem{Horowitz:2009ij}
G.~T. Horowitz and M.~M. Roberts, ``{Zero Temperature Limit of Holographic Superconductors},'' {\em JHEP} {\bf 11} (2009) 015, \href{http://www.arXiv.org/abs/0908.3677}{{\tt 0908.3677}}.

\bibitem{Faulkner:2009wj}
T.~Faulkner, H.~Liu, J.~McGreevy, and D.~Vegh, ``{Emergent quantum criticality, Fermi surfaces, and AdS(2)},'' {\em Phys. Rev. D} {\bf 83} (2011) 125002, \href{http://www.arXiv.org/abs/0907.2694}{{\tt 0907.2694}}.

\bibitem{Albash:2008eh}
T.~Albash and C.~V. Johnson, ``{A Holographic Superconductor in an External Magnetic Field},'' {\em JHEP} {\bf 09} (2008) 121, \href{http://www.arXiv.org/abs/0804.3466}{{\tt 0804.3466}}.

\bibitem{Nicolini:2008aj}
P.~Nicolini, ``{Noncommutative Black Holes, The Final Appeal To Quantum Gravity: A Review},'' {\em Int. J. Mod. Phys. A} {\bf 24} (2009) 1229--1308, \href{http://www.arXiv.org/abs/0807.1939}{{\tt 0807.1939}}.

\bibitem{Ansoldi:2006vg}
S.~Ansoldi, P.~Nicolini, A.~Smailagic, and E.~Spallucci, ``{Noncommutative geometry inspired charged black holes},'' {\em Phys. Lett. B} {\bf 645} (2007) 261--266, \href{http://www.arXiv.org/abs/gr-qc/0612035}{{\tt gr-qc/0612035}}.

\bibitem{Pramanik:2015eka}
S.~Pramanik and S.~Ghosh, ``{AdS-CFT Correspondence in Noncommutative background, related thermodynamics and Holographic Superconductor in Magnetic Field},'' {\em Gen. Rel. Grav.} {\bf 51} (2019), no.~1, 7, \href{http://www.arXiv.org/abs/1509.07825}{{\tt 1509.07825}}.

\bibitem{Pramanik:2014mya}
S.~Pramanik, S.~Das, and S.~Ghosh, ``{Noncommutative extension of AdS{\textendash}CFT and holographic superconductors},'' {\em Phys. Lett. B} {\bf 742} (2015) 266--273, \href{http://www.arXiv.org/abs/1401.7832}{{\tt 1401.7832}}.

\bibitem{Maceda:2019woa}
M.~Maceda and S.~Pati\~no L\'opez, ``{Holographic superconductor from a noncommutative-inspired anti-de Sitter\textendash{}Einstein\textendash{}Born\textendash{}Infeld black hole},'' {\em Int. J. Mod. Phys. D} {\bf 29} (2019), no.~01, 2050003, \href{http://www.arXiv.org/abs/1903.02132}{{\tt 1903.02132}}.

\bibitem{Balasubramanian:1999re}
V.~Balasubramanian and P.~Kraus, ``{A Stress tensor for Anti-de Sitter gravity},'' {\em Commun. Math. Phys.} {\bf 208} (1999) 413--428, \href{http://www.arXiv.org/abs/hep-th/9902121}{{\tt hep-th/9902121}}.

\bibitem{book:17888}
M.~Tinkham, {\em Introduction to superconductivity}.
\newblock International series in pure and applied physics. McGraw Hill, 2nd ed~ed., 1996.

\bibitem{refId0}
{de la Cruz-López, Manuel}, {Herrera-Mendoza, Jhony A.}, {Cartas-Fuentevilla, Roberto}, and {Herrera-Aguilar, Alfredo}, ``Exploring the uv and ir of a type-ii holographic superconductor using a dyonic black hole,'' {\em Eur. Phys. J. Plus} {\bf 139} (2024), no.~9, 786.

\bibitem{10.5555/1098650}
M.~Abramowitz, {\em Handbook of Mathematical Functions, with Formulas, Graphs, and Mathematical Tables}.
\newblock Dover Publications, Inc., USA, 1974.

\bibitem{Olea:2005gb}
R.~Olea, ``{Mass, angular momentum and thermodynamics in four-dimensional Kerr-AdS black holes},'' {\em JHEP} {\bf 06} (2005) 023, \href{http://www.arXiv.org/abs/hep-th/0504233}{{\tt hep-th/0504233}}.

\bibitem{Olea:2006vd}
R.~Olea, ``{Regularization of odd-dimensional AdS gravity: Kounterterms},'' {\em JHEP} {\bf 04} (2007) 073, \href{http://www.arXiv.org/abs/hep-th/0610230}{{\tt hep-th/0610230}}.

\bibitem{PhysRevD.15.2752}
G.~W. Gibbons and S.~W. Hawking, ``Action integrals and partition functions in quantum gravity,'' {\em Phys. Rev. D} {\bf 15} (May, 1977) 2752--2756.

\bibitem{Banerjee:2008du}
R.~Banerjee, B.~R. Majhi, and S.~K. Modak, ``{Noncommutative Schwarzschild Black Hole and Area Law},'' {\em Class. Quant. Grav.} {\bf 26} (2009) 085010, \href{http://www.arXiv.org/abs/0802.2176}{{\tt 0802.2176}}.

\bibitem{Hartnoll:2016apf}
S.~A. Hartnoll, A.~Lucas, and S.~Sachdev, ``{Holographic quantum matter},'' \href{http://www.arXiv.org/abs/1612.07324}{{\tt 1612.07324}}.

\bibitem{Faulkner:2010jy}
T.~Faulkner, H.~Liu, and M.~Rangamani, ``{Integrating out geometry: Holographic Wilsonian RG and the membrane paradigm},'' {\em JHEP} {\bf 08} (2011) 051, \href{http://www.arXiv.org/abs/1010.4036}{{\tt 1010.4036}}.

\bibitem{Faulkner:2010gj}
T.~Faulkner, G.~T. Horowitz, and M.~M. Roberts, ``{Holographic quantum criticality from multi-trace deformations},'' {\em JHEP} {\bf 04} (2011) 051, \href{http://www.arXiv.org/abs/1008.1581}{{\tt 1008.1581}}.

\bibitem{Abrikosov:1956sx}
A.~A. Abrikosov, ``{On the Magnetic properties of superconductors of the second group},'' {\em Sov. Phys. JETP} {\bf 5} (1957) 1174--1182.

\bibitem{Breitenlohner:1982bm}
P.~Breitenlohner and D.~Z. Freedman, ``{Positive Energy in anti-De Sitter Backgrounds and Gauged Extended Supergravity},'' {\em Phys. Lett. B} {\bf 115} (1982) 197--201.

\bibitem{Witten:1998qj}
E.~Witten, ``{Anti-de Sitter space and holography},'' {\em Adv. Theor. Math. Phys.} {\bf 2} (1998) 253--291, \href{http://www.arXiv.org/abs/hep-th/9802150}{{\tt hep-th/9802150}}.

\bibitem{Hartnoll:2007ip}
S.~A. Hartnoll and C.~P. Herzog, ``{Ohm's Law at strong coupling: S duality and the cyclotron resonance},'' {\em Phys. Rev. D} {\bf 76} (2007) 106012, \href{http://www.arXiv.org/abs/0706.3228}{{\tt 0706.3228}}.

\bibitem{Iqbal:2011in}
N.~Iqbal, H.~Liu, and M.~Mezei, ``{Semi-local quantum liquids},'' {\em JHEP} {\bf 04} (2012) 086, \href{http://www.arXiv.org/abs/1105.4621}{{\tt 1105.4621}}.

\bibitem{Hartnoll:2011dm}
S.~A. Hartnoll, D.~M. Hofman, and D.~Vegh, ``{Stellar spectroscopy: Fermions and holographic Lifshitz criticality},'' {\em JHEP} {\bf 08} (2011) 096, \href{http://www.arXiv.org/abs/1105.3197}{{\tt 1105.3197}}.

\bibitem{Cubrovic:2011xm}
M.~Cubrovic, Y.~Liu, K.~Schalm, Y.-W. Sun, and J.~Zaanen, ``{Spectral probes of the holographic Fermi groundstate: dialing between the electron star and AdS Dirac hair},'' {\em Phys. Rev. D} {\bf 84} (2011) 086002, \href{http://www.arXiv.org/abs/1106.1798}{{\tt 1106.1798}}.

\bibitem{Iqbal:2010eh}
N.~Iqbal, H.~Liu, M.~Mezei, and Q.~Si, ``{Quantum phase transitions in holographic models of magnetism and superconductors},'' {\em Phys. Rev. D} {\bf 82} (2010) 045002, \href{http://www.arXiv.org/abs/1003.0010}{{\tt 1003.0010}}.

\bibitem{Laia:2011wf}
J.~N. Laia and D.~Tong, ``{Flowing Between Fermionic Fixed Points},'' {\em JHEP} {\bf 11} (2011) 131, \href{http://www.arXiv.org/abs/1108.2216}{{\tt 1108.2216}}.

\bibitem{Xia:2021jzh}
C.-Y. Xia, H.-B. Zeng, Y.~Tian, C.-M. Chen, and J.~Zaanen, ``{Holographic Abrikosov lattice: Vortex matter from black hole},'' {\em Phys. Rev. D} {\bf 105} (2022), no.~2, L021901, \href{http://www.arXiv.org/abs/2111.07718}{{\tt 2111.07718}}.

\bibitem{Domenech:2010nf}
O.~Domenech, M.~Montull, A.~Pomarol, A.~Salvio, and P.~J. Silva, ``{Emergent Gauge Fields in Holographic Superconductors},'' {\em JHEP} {\bf 08} (2010) 033, \href{http://www.arXiv.org/abs/1005.1776}{{\tt 1005.1776}}.

\bibitem{Natsuume:2024ril}
M.~Natsuume, ``{What is the dual Ginzburg-Landau theory for holographic superconductors?},'' \href{http://www.arXiv.org/abs/2407.13956}{{\tt 2407.13956}}.

\bibitem{Banerjee:2009xx}
R.~Banerjee, S.~Gangopadhyay, and S.~K. Modak, ``{Voros product, Noncommutative Schwarzschild Black Hole and Corrected Area Law},'' {\em Phys. Lett. B} {\bf 686} (2010) 181--187, \href{http://www.arXiv.org/abs/0911.2123}{{\tt 0911.2123}}.

\bibitem{Seiberg:1999vs}
N.~Seiberg and E.~Witten, ``{String theory and noncommutative geometry},'' {\em JHEP} {\bf 09} (1999) 032, \href{http://www.arXiv.org/abs/hep-th/9908142}{{\tt hep-th/9908142}}.

\bibitem{Martinez-Carbajal:2021kfp}
D.~Mart\'\i{}nez-Carbajal, M.~de~la Cruz, S.~Pati\~no L\'opez, and L.~D. Herrera-Z\'u\~niga, ``{Implications of Seiberg\textendash{}Witten map on Type-I superconductors},'' {\em Int. J. Mod. Phys. A} {\bf 37} (2022), no.~35, 2250218, \href{http://www.arXiv.org/abs/2110.08469}{{\tt 2110.08469}}.

\end{thebibliography}\endgroup
\bibliographystyle{./utphys}

\end{document}